\documentclass[twocolumn,epjc3]{svjour3}  

\pdfoutput=1

\smartqed  
\RequirePackage{graphicx}

\usepackage{hyperref}
\usepackage{widetext}
\usepackage{subfig}
\usepackage{graphicx}
\usepackage{amsmath}
\usepackage{xspace}
\usepackage{xcolor}
\usepackage{float}
\usepackage{soul}
\usepackage{hyphenat}
\newcommand{\eq}[1]{eq.~\eqref{eq:#1}}

\renewcommand{\sec}[1]{sec.~\ref{sec:#1}}

\newcommand{\subsec}[1]{sec.~\ref{subsec:#1}}
\newcommand{\fig}[1]{fig.~\ref{fig:#1}}
\newcommand{\Fig}[1]{Fig.~\ref{fig:#1}}

\newcommand{\ord}[1]{\mathcal{O}(#1)}

\newcommand{\df}{\mathrm{d}}

\newcommand{\as}{\alpha_{\rm s}}

\newcommand{\Tau}{\mathcal{T}}

\newcommand{\GeV}{\,\mathrm{GeV}}

\newcommand{\nn}{\nonumber}

\newcommand{\cP}{\mathcal{P}}

\newcommand{\cut}{\mathrm{cut}}

\newcommand{\NLO}{\mathrm{NLO}}
\newcommand{\NNLO}{\mathrm{NNLO}}

\newcommand{\NLL}{\mathrm{NLL}}
\newcommand{\NNLL}{\mathrm{NNLL}}

\newcommand{\one}{{(1)}}

\newcommand{\lqcd}{\Lambda_\mathrm{QCD}}
\newcommand{\dsigMC}{\df\sigma^\textsc{mc}}

\newcommand{\geneva}{\textsc{Geneva}\xspace}

\newcommand{\pythia}{\textsc{Pythia}\xspace}
\newcommand{\pythiaEight}{\textsc{Pythia8}\xspace}

\newcommand{\atlas}{\mbox{ATLAS}\xspace}
\newcommand{\cms}{\mbox{CMS}\xspace}

\def\pt{\ensuremath{p_\mathrm{T}}\xspace}

\newcommand{\Nchg}{\ensuremath{N_\text{ch}}\xspace}
\newcommand{\ZpT}{\ensuremath{p_\mathrm{{T}}^{Z}}\xspace}

\newcommand{\dNchgdetadphi}{\ensuremath {N_\text{ch}/\delta\eta\,\delta\phi} \xspace}
\newcommand{\dpTsumdetadphi}{\ensuremath{\sum\!p_T/\delta\eta\,\delta\phi }\xspace}
\newcommand{\Nchgdetadphi}{\ensuremath {N_\text{ch}/\delta\eta\,\delta\phi} \xspace}
\newcommand{\pTsumdetadphi}{\ensuremath{\sum\!p_T/\delta\eta\,\delta\phi}\xspace}
\newcommand{\ptmean}{\ensuremath{\langle\pt\rangle}\xspace}

\renewcommand{\thesection}{\arabic{section}}

\allowdisplaybreaks[2]

\journalname{Eur. Phys. J. C}

\begin{document}


\title{Underlying event sensitive observables in Drell-Yan production using GENEVA}

\author{Simone Alioli\thanksref{e1,addr1} \and Christian W.~Bauer\thanksref{e2,addr2} \and Sam Guns\thanksref{e3,addr2} \and Frank J.~Tackmann\thanksref{e4,addr3}}

\institute{ CERN Theory Division, CH-1211, Geneva 23, Switzerland \label{addr1} \and Ernest Orlando Lawrence Berkeley National Laboratory, University of California, Berkeley, CA 94720, U.S.A.\label{addr2} \and Theory Group, Deutsches Elektronen-Synchrotron (DESY), D-22607 Hamburg, Germany  \label{addr3}}

\thankstext{e1}{e-mail: simone.alioli@cern.ch}
\thankstext{e2}{e-mail: cwbauer@lbl.gov}
\thankstext{e3}{e-mail: sguns@lbl.gov}
\thankstext{e4}{e-mail: frank.tackmann@desy.de}

\date{May 23, 2016}


\maketitle
 
\begin{abstract}
We present an extension of the \geneva Monte Carlo framework to include multiple parton interactions (MPI) provided by \pythiaEight.
This allows us to obtain predictions for underlying-event sensitive measurements in Drell-Yan production,
in conjunction with \geneva's fully-differential $\NNLO$ calculation, $\NNLL'$ resummation for the 0-jet resolution variable (beam thrust), and $\NLL$ resummation for the $1$-jet resolution variable.
We describe the interface with the parton shower algorithm and MPI model of \pythiaEight, which preserves both the precision of partonic $N$-jet cross sections in \geneva as well as the shower accuracy and good description of soft hadronic physics of \pythiaEight.
We present results for several underlying-event sensitive observables and compare to data from \atlas and \cms as well as to standalone \pythiaEight predictions. This includes a comparison with the recent \atlas measurement of the beam thrust spectrum, which provides a potential avenue to fully disentangle the physical effects from the primary hard interaction, primary soft radiation, multiple parton interactions, and nonperturbative hadronization.
\end{abstract}


\section{Introduction}
\label{sec:intro}

Exclusive Monte-Carlo event generators are an important tool to make theoretical predictions for collider observables. By including both perturbative and nonperturbative effects, exclusive generators are able to provide predictions for a wide range of observables, whether they are dominated by short-distance physics or not. They aim to correctly describe physical effects over a wide range of energy scales, including
\begin{enumerate}
\item  The perturbative effects of the primary hard interaction,
\item  The perturbative evolution of the products of the primary hard interaction,
\item  Additional interactions between partons within the same proton (MPI)
\item  Nonperturbative physics such as hadronization, beam remnants, transverse momenta in the colliding partons, etc.
\end{enumerate}

For sufficiently inclusive observables, it is enough to include only the perturbative effects of the primary hard interaction to achieve precise predictions, and for such observables one can directly compare measured data against partonic calculations. For less inclusive, resummation-sensitive observables, the perturbative evolution of the primary interaction becomes important, while MPI and most of the nonperturbative effects still give a subdominant contribution. Finally, there is a large class of exclusive observables for which all physical effects mentioned above contribute to similar extent.

The concept of the underlying event (UE)~\cite{Akesson:1984yp,Sjostrand:1987su,Marchesini:1988hj,Affolder:2001xt} was introduced as a means to describe the soft hadronic activity that underlies the primary hard interaction. Typically, the effects arising from MPI are directly associated with the UE, and
the traditional approach taken to study the UE is to define observables that have strong sensitivity to MPI and for which the effect from the primary interaction is reduced. In the following, we call such observables UE-sensitive observables.
However, in principle all of the above contributions can give rise to the effects that are experimentally associated with soft activity and the underlying event, which makes a precise theoretical definition of the UE  challenging.
For example, it is well known that including higher-order perturbative corrections to the primary interaction in parton-shower Monte Carlos can give a nontrivial contribution to traditional UE measurements~\cite{Cacciari:2009dp,Chatrchyan:2012tb}. Similarly, interference effects due to perturbative soft initial-state radiation in the primary interaction can contribute to many observables in a similar way than MPI effects~\cite{Stewart:2014nna}.

UE-sensitive observables are typically constructed by dividing each event into distinct angular regions~\cite{Affolder:2001xt}. The ``toward'' and ``away'' regions are defined to be aligned with the directions of the products of the primary hard $2 \to 2$ interaction, while the ``transverse`` region is the complementary coverage of the solid angle. While the toward and away regions are typically dominated by the primary interaction, those effects are reduced in the transverse region. This makes observables measured in the transverse region sensitive to MPI. However, the effects from the primary interaction, in particular soft radiation and hadronization effects, give a sizable contribution in the transverse regions and need to be included in a proper description.

In recent years, there has been much effort to combine precision calculations of the hard interaction with fully exclusive Monte Carlo generators, which aim to describe the perturbative evolution of the primary partons and add nonperturbative physics and MPI effects through physical models. This started with the combination of leading order (LO) predictions for several multiplicities with parton showers~\cite{Catani:2001cc,Lonnblad:2001iq}, and shortly thereafter methods to combine next-to-leading order (NLO) calculations were developed~\cite{Frixione:2002ik,Frixione:2003ei,Nason:2004rx,Frixione:2007vw,Alioli:2010xd}. By now there are several methods available to combine multiple NLO calculations with parton showers~\cite{Alioli:2011nr,Hoeche:2012yf,Gehrmann:2012yg,Frederix:2012ps,Platzer:2012bs,Alioli:2012fc,Lonnblad:2012ix,Hamilton:2012rf,Luisoni:2013cuh}. More recently, combinations of specific Drell-Yan like NNLO calculations with parton showers have been presented in Refs. \cite{Hamilton:2013fea,Hoeche:2014aia,Karlberg:2014qua,Hoche:2014dla,Hamilton:2015nsa,Alioli:2013hqa,Alioli:2015toa}.

As the main focus of these theoretical developments is to increase the perturbative precision of the primary interaction, most studies of this focus on observables that are sensitive to primary perturbative effects (1. and 2. above) but are insensitive to MPI. Since the purpose of these methods is to combine the perturbative calculations with the versatility of fully exclusive  Monte-Carlo generators, it is important to also study how well observables sensitive to soft hadronic activity and underlying event are described in these approaches, but so far there have been only few studies discussing the impact on UE-sensitive observables in detail~\cite{Lonnblad:2012ix,Campbell:2013vha,Frederix:2015eii,Blok:2015afa}.

In this paper, we perform such a study for the event generator \geneva~\cite{Alioli:2012fc,Alioli:2015toa} interfaced to \pythiaEight~\cite{Sjostrand:2006za,Sjostrand:2007gs,Sjostrand:2014zea}, focusing on Drell-Yan neutral-current production.
By comparing to UE-sensitive observables, the predictions depend on the perturbative calculations in \geneva, as well as the perturbative parton shower, the hadronization model, and the MPI model implemented in \pythiaEight. Of those observables, some are mostly independent of the hard interaction, while others contain sensitivity to both the long distance physics as well as the hard process (for example when long distance observables are shown in bins of the transverse momentum of the $Z$ boson).  We will show that predictions for observables that are mostly independent of the hard interaction, \geneva{}+\pythiaEight yields results that are very similar to those obtained by running \pythiaEight directly. This indicates that the constrained shower used in \geneva is not spoiling the accuracy of the parton shower, or the model used to describe MPI and nonperturbative effects. On the other hand, observables that have sensitivity to the hard kinematics are improved compared to \pythiaEight standalone, as expected.

As already mentioned, while UE-sensitive observables are designed to enhance the effects of MPI, they still are very sensitive to the perturbative soft radiation from the primary interaction, and it is typically difficult to fully disentangle these two effects. An alternative approach~\cite{Stewart:2014nna} is to consider observables for which the perturbative (and in principle also nonperturbative) soft effects from the primary interactions can be explicitly accounted for with field-theoretic methods and calculated to high precision. This then allows one to more directly isolate the physical effects due to MPI and test its modelling. Beam thrust~\cite{Stewart:2009yx} (or $0$-jettiness) provides such an observable. In \geneva, beam thrust is used as jet resolution variable and is resummed to $\NNLL'$ accuracy, providing a precise description of the physics originating from the primary interaction. Thus, comparing predictions from \geneva interfaced to the hadronization and MPI model of \pythia with experimental data for this observable provides one of the best ways to isolate the effects from MPI.

This paper is organized as follows: In \sec{ShowerInterface} we review the perturbative inputs in \geneva and how \geneva is interfaced with \pythiaEight. In \sec{ComparisonWithData} we compare the predictions of \geneva{}+\pythiaEight with Drell-Yan measurements from \atlas and \cms of several UE-sensitive observables. In \sec{Beamthrust} we present a comparison of \geneva{}+\pythiaEight and data from \atlas for beam thrust. Our conclusions are presented in \sec{conclusions}.

\section{Review of \geneva and its \pythiaEight interface}
\label{sec:ShowerInterface}

In this section, we discuss the theoretical setup of \geneva. We first review the partonic calculations used to obtain the $\NNLL'+$NNLO perturbative accuracy, and then the interface with the parton shower. We present here only the main results, and for a more detailed presentation of the method we refer the reader to refs.~\cite{Alioli:2012fc,Alioli:2013hqa,Alioli:2015toa}.

\subsection{The partonic calculation}
\label{subsec:PartonicCalculation}
As discussed in detail in ref.~\cite{Alioli:2015toa}, \geneva separates the available partonic phase space into exclusive 0/1-jet and an inclusive 2-jet cross section:
\begin{align} \label{eq:NNLOevents}
\text{$\Phi_0$ events: }
& \qquad \frac{\dsigMC_0}{\df\Phi_0}(\Tau_0^\cut)
\,,\nn \\
\text{$\Phi_1$ events: }
& \qquad
\frac{\dsigMC_{1}}{\df\Phi_{1}}(\Tau_0 > \Tau_0^\cut; \Tau_{1}^\cut)
\,, \nn\\
\text{$\Phi_2$ events: }
& \qquad
\frac{\dsigMC_{\ge 2}}{\df\Phi_{2}}(\Tau_0 > \Tau_0^\cut, \Tau_{1} > \Tau_{1}^\cut)
\,,\end{align}
each of which is fully differential in their jet phase space. Jet resolution variables $\Tau_0$ and $\Tau_1$ are used to separate the exclusive 0-jet from the inclusive 1-jet, and the exclusive 1-jet from the inclusive 2-jet cross section, respectively. By including the resummation of the $\Tau_0^\cut$ and $\Tau_1^\cut$ dependence to high order, the numerical values of these two jet resolution parameters can be chosen to be small, such that the phase space is dominated by the inclusive 2-jet sample.

The 0-jet cross section is given by
\begin{align} \label{eq:0master}
\frac{\dsigMC_0}{\df\Phi_0}(\Tau_0^\cut)
&= \frac{\df\sigma^{\rm NNLL'}}{\df\Phi_0}(\Tau_0^\cut)
+  \frac{\df\sigma_0^{{\rm NNLO_0}}}{\df\Phi_{0}}(\Tau_0^\cut)
 \nn \\ & \quad
- \biggl[\frac{\df\sigma_0^{\rm NNLL'}}{\df\Phi_{0}}(\Tau_0^\cut) \biggr]_{\rm NNLO_0}\,.
\end{align}
It is correct to NNLO at fixed order, and to $\NNLL'$ in the resummation of the $\Tau_0^\cut$ dependence. The  $[ \ldots ]_{\rm NNLO_0}$ notation indicates that the result inside the brackets is expanded up to ${\cal O}(\as^2)$ relative accuracy.

The 1-jet cross section is
\begin{widetext}
\begin{align} \label{eq:1master}
\frac{\dsigMC_{1}}{\df\Phi_{1}} (\Tau_0 > \Tau_0^\cut; \Tau_{1}^\cut)
&= \nn \\ & \hspace{-2.5cm}\frac{\df\sigma_{\geq 1}^C}{\df\Phi_1} \, U_1(\Phi_1, \Tau_1^\cut)
- B_1(\Phi_1)\, U_1^\one(\Phi_1, \Tau_1^\cut)
+ \int\!\df \Phi^\Tau_{\textrm{rad}} \,  B_{2}(\Phi_1,\Phi^\Tau_{\textrm{rad}})\, \theta(\Tau_{1} < \Tau_1^\cut)
- \df \Phi^{\rm C}_{\textrm{rad}} C_{2}(\Phi_{1},\Phi^{\rm C}_{\textrm{rad}})
\,,\end{align}
\end{widetext}

where $U_1(\Phi_1, \Tau_1^\cut)$ describes the resummation of the $\Tau_1^\cut$ dependence, and $U_1^{(1)}(\Phi_1, \Tau_1^\cut)$  the first order of its expansion in $\alpha_s$. The exclusive 1-jet cross section needs to include the $\Phi_2$ contributions below the $\Tau_1^\cut$. To perform this integration one writes the integral over $\Phi_2$ as $\df  \Phi_2~\equiv~\df \Phi_1 \df \Phi_{\textrm{rad}}$, where $\Phi_{\textrm{rad}}$ denotes the 3 additional radiation variables to define $\Phi_2$,  and $\df \Phi_{\textrm{rad}}$ includes the appropriate Jacobian factor.
This requires the introduction of a phase space map $\Phi_2 \equiv \Phi_2(\Phi_1,\Phi_{\textrm{rad}})$ for both the real-emission and the subtraction contribution. The two maps need not be the same, but they need to agree in the collinear and soft limits for the subtraction to pointwise cancel the IR singularities. We have labeled the two maps as $\Phi_2(\Phi_1,\Phi^\Tau_{\textrm{rad}})$ and $\Phi_2(\Phi_1,\Phi^{\rm C}_{\textrm{rad}})$, respectively. The real-emission and the subtraction contributions are then directly expressed as function of these maps as $B_2(\Phi_1,\Phi^\Tau_{\textrm{rad}})$ and  $C_2(\Phi_1,\Phi^{\rm C}_{\textrm{rad}})$.
It is important that the map $\Phi_{2}(\Phi_1,\Phi^\Tau_{\textrm{rad}})$  preserves the value of $\Tau_0$, i.e. $\Tau_0 [\Phi_{2}(\Phi_1,\Phi^\Tau_{\textrm{rad}})]= \Tau_0 (\Phi_{1})$ in order for the correct $\Tau_0$-singular structure of ${\frac{\dsigMC_{1}}{\df\Phi_{1}} (\Tau_0 > \Tau_0^\cut; \Tau_{1}^\cut)}$ to be reproduced at $\NNLL'$. This is discussed in detail in  ref.~\cite{Alioli:2015toa}. The map used for the subtraction is the standard FKS map~\cite{Frixione:2007vw}.

The expression for $\df\sigma^{C}_{\geq 1} / \df\Phi_{1}$ is given by
\begin{align}
\frac{\df\sigma^{C}_{\geq 1}}{\df\Phi_{1}}
&= \frac{\df\sigma^{\rm NNLL'}}{\df\Phi_{0}\df \Tau_0} \, \cP(\Phi_1)
+ (B_1 + V_1^C)(\Phi_1) \nn \\ & \quad
- \biggl[\frac{\df\sigma^{\rm NNLL'}}{\df\Phi_{0}\df \Tau_0}\,\biggr]_{\NLO_1} \, \cP(\Phi_1)
\,,\end{align}
where the normalized splitting function $P(\Phi_1)$ described in~\cite{Alioli:2015toa} makes the $\Tau_0$ resummation differential in the full $\Phi_1$ phase space. Note that we have omitted the $\theta$-functions enforcing the constraint $\Tau_0 > \Tau_0^\cut$ for notational simplicity. 

The inclusive 2-jet cross section is given by
\begin{widetext}
\begin{align} \label{eq:2master}
\frac{\dsigMC_{\geq 2}}{\df\Phi_{2}} (\Tau_0 > \Tau_0^\cut, \Tau_{1}>\Tau_{1}^\cut)
& = \nn\\ &\hspace{-2cm} \frac{\df\sigma_{\geq 1}^C}{\df\Phi_1^\Tau}\, U_1'(\Phi_1^\Tau, \Tau_1) \, \cP(\Phi_2) 
- B_1(\Phi_1^\Tau)\,U_1^{\one\prime}(\Phi_1^\Tau, \Tau_1)\,\cP(\Phi_2)
+ B_2(\Phi_2)\,[1 - \Theta^\Tau(\Phi_2)\,\theta(\Tau_1 < \Tau_1^\cut)]
\,,\end{align}
\end{widetext}
where $U_1'(\Phi_1, \Tau_1)$ denotes the derivative of $U_1(\Phi_1, \Tau_1)$ with respect to  $\Tau_1$, and we have again omitted the $\theta$-functions enforcing the constraint $\Tau_0 > \Tau_0^\cut$ and $\Tau_1 > \Tau_1^\cut$. 
The dependence on the map used for the projection $\Phi_{1}^\Tau \equiv \Phi_1^\Tau(\Phi_2)$ is also kept manifest here. This must correspond to the inversion of the mapping from $\Phi_{1}\to\Phi_{2}$ used in \eq{1master} for the real-emission contribution. 
However, such a map does not necessarily cover all of $\Phi_2$ phase space, as it can omit nonsingular regions. Therefore, the regions in $\Phi_2$ with $\Tau_1 < \Tau_1^\cut$ which are not covered by the phase space map  have to be included in the inclusive 2-jet cross section above
by means of the $\Theta^\Tau$ function.
 
Finally, we point out one important difference of the implementation used here compared to ref.~\cite{Alioli:2015toa}. In that paper, the $\Tau_1$ resummation was performed to LL accuracy. While this was correct to the order stated, it left a logarithmic dependence on $\Tau_1$ in the nonsingular contributions to \eq{2master}. This dependence can be removed by performing the  $\Tau_1$ resummation to higher accuracy. For the results of this paper we have implemented the resummation of $\Tau_1$ to full NLL accuracy.  Including the full kinematic dependence, the relevant NLL evolution factor is given by~\cite{Stewart:2010tn,Pietrulewicz:2016nwo}
\begin{widetext}
\begin{align}
 U_1(\Phi_1, \Tau_1^\cut) & = \frac{U}{\Gamma\bigg(1+2(2 C_F + C_A) \left[\eta_\Gamma^{\rm NLL} \left(\mu_S,\mu_H\right)-\eta_\Gamma^{\rm NLL} \left(\mu_J,\mu_H\right)\right]\bigg)} \qquad \rm with
\end{align}

\begin{align}
\ln U &=  2 (2 C_F + C_A)\Bigg[ 
2 K_\Gamma^{\rm NLL}\left(\mu_J, \mu_H\right)  -  K_\Gamma^{\rm NLL}\left(\mu_S,\mu_H\right) \Bigg]
 + 
2 C_F \Bigg[ - \eta_\Gamma^{\rm NLL}\left(\mu_J, \mu_H\right) \ln \left(\frac{w_q w_{\bar q}}{\mu_H^2}\right)  \\
& \qquad + \eta_\Gamma^{\rm NLL} \left(\mu_S,\mu_H\right)\ln \left(\frac{w_q w_{\bar q}}{s_{q\bar q}}\right)\Bigg]
 + 
C_A \Bigg[ - \eta_\Gamma^{\rm NLL} \left(\mu_J,\mu_H\right) \ln \left(\frac{w_g^2}{\mu_H^2}\right) + \eta_\Gamma^{\rm NLL} \left(\mu_S,\mu_H\right) \ln \left(\frac{w_g^2 s_{q \bar q}}{s_{qg}s_{\bar qg}}\right)\Bigg]
 \nn \\
& \qquad + 
K_\gamma^{\rm NLL}\left(\mu_J, \mu_H\right) - 2 \gamma_{\rm E} (2 C_F + C_A)\left[\eta_\Gamma^{\rm NLL} \left(\mu_S,\mu_H\right)-\eta_\Gamma^{\rm NLL} \left(\mu_J,\mu_H\right)\right] \nn
\,.\end{align}
\end{widetext}
Here we have defined the functions
\begin{align}
K_\Gamma^{\rm NLL}(\mu_1, \mu_2) & = -\frac{\Gamma_0}{4\beta_0^2} \biggl[ \frac{4\pi}{\alpha_s(\mu_1)}\left( 1 - \frac{1}{r} - \ln r\right) \nn \\ &+ \left(\frac{\Gamma_1}{\Gamma_0} - \frac{\beta_1}{\beta_0}\right)\left( 1 - r + \ln r\right) + \frac{\beta_1}{2\beta_0}  \ln^2 r \biggr]
\nn\\
\eta_\Gamma^{\rm NLL}(\mu_1, \mu_2) &= -\frac{1}{2} \frac{\Gamma_0}{\beta_0} \biggl[ \ln r  + \frac{\alpha_s(\mu_1)}{4\pi} \nn \\ &\qquad \qquad \qquad  \quad \times \left(\frac{\Gamma_1}{\Gamma_0} - \frac{\beta_1}{\beta_0}\right)(r-1)\biggr]
\nn\\
K_\gamma^{\rm NLL}(\mu_1, \mu_2) &= -\frac{1}{2} \frac{\gamma_0}{\beta_0} \ln r
\,,\end{align}
with $r = \alpha_s(\mu_2) / \alpha_s(\mu_1)$ and the dependence on $\Tau_1^\cut$ is through the dependence on the scales
\begin{align}
\mu_S = \Tau_1^\cut\,, \qquad \mu_H = \Tau_1^{\max}\,, \qquad \mu_J^2 = \mu_S \, \mu_H
\,.\end{align}
Here, $\Tau_1^{\max}$ is the value at which the $\Tau_1$ resummation is turned off, which is chosen near the maximum kinematically allowed value of $\Tau_1$ for a given phase space point $\Phi_1$.
The cusp and noncusp anomalous dimensions entering above are given by
\begin{align}
\Gamma_0 & = 4
\,,\qquad
\Gamma_1 = 4\biggl[\Bigr(\frac{67}{9} - \frac{\pi^2}{3}\Bigl) C_A - \frac{20}{9} T_F n_f \biggr]
\,,\nn\\
\gamma_0 & = 12 C_F + 2 \beta_0
\,, \qquad
\beta_0 = \frac{11}{3} C_A - \frac{4}{3} T_F n_f
\,.\end{align}
The kinematical terms are
\begin{align}
& s_{ab} = p_a^- p_b^+\,, \quad  s_{a1} = p_a^- p_1^+\,, \quad s_{b1} = p_a^+ p_1^-
\\ \nn
& w_a = p_a^- e^{-Y_V}\,, \quad  w_b = p_b^+ e^{Y_V}\,, \quad w_1 = p_1^+ e^{Y_V} + p_1^- e^{-Y_V}
\,,\end{align}
where $p_a$, $p_b$, and $p_1$ are the massless four-momenta of the $\Phi_1$ phase space point, and ${p^+ = p^0 - p^3}$,  ${p^- = p^0 + p^3}$. The assignment of $p_a$, $p_b$, and $p_1$ to $p_q$, $p_{\bar q}$, and $p_g$ is according to the flavor structure of $\Phi_1$. For example, for a $q \bar q \to Z g$ flavor structure we have $p_q = p_a$, $p_{\bar q} = p_b$ and $p_g = p_1$.

\subsection{Interface to the parton shower}
\label{subsec:ShowerInterface}

When running a parton shower on the partonic events generated by \geneva, one recovers the emissions that were integrated over in the definitions of the jet cross-sections. This includes the integrations over the real emission below $\Tau_0^\cut$ and $\Tau_1^\cut$ for the exclusive 0- and 1-jet cross sections, and also the integrations over higher multiplicities that are included through the resummation in the inclusive 2-jet rate. The parton shower must act in such a way that it does not affect the integrated jet cross-sections, while at the same time giving fully exclusive final states with a large multiplicity of particles.
The difficulty is in telling the parton shower how to cover precisely those regions of the phase space which were integrated over. In particular, the phase space map $\Phi_1 \to \Phi_2(\Phi_1, \Phi^\Tau_{\rm rad})$ results in very complicated integration hyper-surfaces when written in terms of the parton shower variables.

In ref.~\cite{Alioli:2012fc}, where $e^+ e^- \to {\rm jets}$ was considered, these constraints were implemented by explicitly restricting the kinematics of the showered event. The approach taken in ref.~\cite{Alioli:2015toa} improved on this. Since the precise definitions of the phase space maps are only required for the multiplicities covered by the perturbative calculation (up to 2 extra partons for  the case considered), one can perform the emissions up to $\Phi_2$ using analytical resummation that is based on the exact phase space maps used, and only let the parton shower handle the emissions after that point.
Thus, one analytically adds the first emission off a $\Phi_1$ event, down to the scale $\Lambda_1$, using the $\Tau_1$ resummation and phase space map contained in \geneva, giving
\begin{align}
\frac{\dsigMC_{1}}{\df\Phi_{1}} (\Tau_0 > \Lambda_0, \Tau_0^\cut, \Tau_1^\cut, \Lambda_1)
\label{eq:analyticaShower1}
&= \\ \nn &\hspace{-3cm} \frac{\df\sigma_{1}}{\df\Phi_{1}} (\Tau_0 > \Tau_0^\cut, \Tau_1^\cut) \, U_1(\Tau_1^\cut, \Lambda_1)
\,,\\
\frac{\dsigMC_{\geq 2}}{\df\Phi_{2}} (\Tau_0 > \Lambda_0, \Tau_{1}>\Lambda_1, \Tau_0^\cut, \Tau_1^\cut)
&= \label{eq:analyticaShower2} \\ \nn &\hspace{-3cm} \frac{\dsigMC_{\geq 2}}{\df\Phi_{2}} (\Tau_0 > \Tau_0^\cut, \Tau_{1}>\Tau_{1}^\cut) 
\\
& \hspace{-3cm}
+ \frac{\df}{\df \Tau_1} \, \frac{\dsigMC_{1}}{\df\Phi_{1}} (\Tau_0 > \Lambda_0, \Tau_0^\cut, \Tau_1^\cut, \Tau_1)  \nn \\ & \hspace{-2cm} \nn \times \cP(\Phi_2) 
\theta(\Tau_1^{\max} > \Tau_1 > \Lambda_1)
\,.\end{align}

Choosing $\Lambda_1 \sim \lqcd$, such that the Sudakov factor $ U_1(\Tau_1^\cut, \Lambda_1)$ becomes very small, results in a vanishing $\frac{\dsigMC_{1}}{\df\Phi_{1}} (\Tau_0 > \Lambda_0, \Tau_0^\cut, \Tau_1^\cut, \Lambda_1) $ contribution.

The remaining constraint on the shower is now much simpler, namely $\Tau_2(\Phi_N) \le \Tau_1(\Phi_2)$, and no constraint on $\Tau_0(\Phi_N)$ and $\Tau_1(\Phi_N)$ is applied. 
This enforces that the subsequent emissions do not give the dominant contribution to $\Tau_1$. The $\Tau_2(\Phi_N) \le \Tau_1(\Phi_2)$ constraint is implemented through a veto, which retries the shower until the resulting event satisfies it. This effectively sets the starting scale of the shower to $\Tau_1(\Phi_2)$, essentially mimicking a $\Tau_M$-ordered shower for $M \le 2$.
It was shown explicitly in ref.~\cite{Alioli:2015toa} that this implementation retains the perturbative accuracy of the partonic calculation, including the NNLL$'+$NNLO accuracy of the $\Tau_0$ distribution.

\subsection{Interface to multiple parton interactions}
\label{subsec:ShowerMPI}

In ref.~\cite{Alioli:2015toa}, the focus has been on the description of the primary interaction and correspondingly on observables that did not have sensitivity to MPI. For this reason, MPI was turned off when interfacing to \pythia. Adding a model of MPI to the constrained shower requires some care.

The factorization formula underlying the perturbative description of the Drell-Yan beam thrust spectrum in \geneva~\cite{Stewart:2009yx,Stewart:2010pd}
\begin{align}
\label{eq:Tau0Factorization}
\frac{\df \sigma^{\rm SCET}}{\df \Phi_0 \df \Tau_0}
&=
\sum_{ij} \frac{\df\sigma_{ij}^B}{\df\Phi_0} H_{ij} (Q^2, \mu) \int\!\df t_a\, \df t_b \, B_i (t_a, x_a, \mu)
\nn \\ & \qquad \times
B_j (t_b, x_b, \mu) \, S\Bigl(\Tau_0 - \frac{t_a + t_b}{Q}, \mu \Bigr)
\,,\end{align}
describes the primary hard interaction as well as its perturbative evolution. The beam functions describe the effects of perturbative collinear initial-state radiation (ISR) to $\ord{\as^2}$~\cite{Stewart:2010qs,Gaunt:2014xga}.
The soft function $S$ describes soft ISR and contains contributions due to interference between soft ISR from the two colliding partons, which contributes to UE-sensitive observables. On the other hand, MPI effects are not captured by \eq{Tau0Factorization}~\cite{Gaunt:2014ska,Rothstein:2016bsq}.

The jet resolution variables $\Tau_N$ chosen in \geneva are sensitive to MPI effects. This is because $N-$jettiness is a global observable, such that every particle in the final state contributes, whether it arises from a primary or secondary partonic interaction. Since only effects from the primary interaction are included in the perturbative description of \geneva, the effects from MPI are unconstrained. Therefore, the kinematical constraint  $\Tau_2(\Phi_N) \le \Tau_1(\Phi_2)$ should only be applied on the showering of the primary interaction, and should not include the effects of MPI. The simplest way would be to first add the parton shower to the partons from the primary interaction as produced by \geneva, applying the shower constraints discussed above. After that, MPI effects can be included without any constraints.

In the \pythiaEight Monte Carlo, however, the primary shower and the MPI effects are interleaved~\cite{Sjostrand:2004ef}. The solution we adopt is to separate the final state particles into those that arise from the showering of the primary interaction, and those that arise from MPI. This can be achieved using the event record in \pythiaEight, which keeps track of the origins of each particle in the final state%
\footnote{Note that this separation only makes sense if rescattering effects~\cite{Corke:2009tk}, which allow the scattering of particles from the primary interaction with particles arising from MPI, is turned off.}.
Upon checking if the showered event passes the kinematical constraints, we only use final state particles originating from the primary interaction in the calculation of $\Tau_0$ and $\Tau_1$.
The MPI model included in the parton shower is therefore allowed to violate the kinematic constraints, while the perturbative part of the shower is not.

As discussed above, MPI contributions are not properly captured by \eq{Tau0Factorization}. For this reason, we currently rely entirely on the MPI model in \pythia~\cite{Sjostrand:1987su,Sjostrand:2004pf,Sjostrand:2004ef} to describe the corresponding physics.
Over the past few years, there has been significant progress in the proper field-theoretic description of the associated effects, see e.g.~\cite{Blok:2010ge,Blok:2011bu,Blok:2012mw,Blok:2013bpa,Diehl:2011yj,Manohar:2012jr,Manohar:2012pe,Kasemets:2012pr,Gaunt:2014ska,Diehl:2015bca,Rothstein:2016bsq}, and in the future, one can envision the extension of the perturbative calculations used in \geneva to include such effects, at least in part, which would then directly constrain the MPI model.

\section{Comparison with \atlas and \cms data}
\label{sec:ComparisonWithData}
Both \atlas~\cite{Aad:2014jgf} and \cms~\cite{Chatrchyan:2012tb}  have measured observables sensitive to the UE activity in Drell-Yan events. In this section we compare \geneva predictions to these measurements. We also include the results from a standalone \pythiaEight run with the same tune, to investigate if and how adding the $\NNLL'+$NNLO perturbative effects of \geneva to \pythiaEight changes those results~\footnote{We remind the reader that standalone \pythiaEight results for Drell-Yan production by default include a first-order matching to the leadind-order matrix elements that allows to populate the hard jet regions.} We expect that for observables that  are not sensitive to hard interactions, \geneva{}+\pythiaEight is close to the results of \pythiaEight alone, showing that the implementation discussed in \sec{ShowerInterface} does not spoil the logarithmic structure of the shower. For observables which are sensitive to the hard interaction, we expect that the more accurate perturbative information in \geneva gives an improvement in the description of the data. 

For the results presented in the following we have run \geneva for Drell-Yan production $pp\to Z/\gamma^* \to \mu^+ \mu^-$  at a center-of-mass energy of $7$~TeV, using the central set of the PDF4LHC15~\cite{Butterworth:2015oua} NNLO parton distribution functions from \textsc{Lhapdf6}~\cite{Buckley:2014ana}.
All the  matrix elements, including color and spin-correlations, are obtained from \textsc{OpenLoops}~\cite{Cascioli:2011va}, which improves speed and stability compared to the previous implementation. The invariant mass for the muon pair has been restricted to the range $60 < m_{\mu^+\mu^-} < 120$ GeV.
The other parameters relevant for our calculation are
\begin{eqnarray}
&&M_Z =  91.1876~{ \rm GeV},\quad \Gamma_Z =  2.4952~{\rm GeV}, 
\\&& \sin^2 \theta_W^{\rm eff} =  0.2226459 ,\quad   \alpha_{\rm em }^{-1}(M_Z) = 132.338\,, \nn
\\&& \Tau_0^\cut = 1 \, {\rm GeV}\,, \quad \Tau_1^\cut = 1 \, {\rm GeV} \nn
\,.\end{eqnarray}
The renormalization and factorization scales in the fixed-order contribution and the hard scale in the resummation part are taken equal to the dilepton invariant mass. Both are varied by a factor of 2 in either direction to estimate the fixed-order uncertainties. The resummation uncertainties are evaluated using appropriate profile scale variations~\cite{Ligeti:2008ac,Abbate:2010xh,Gangal:2014qda}, and the total perturbative uncertainties are obtained according to the procedure presented in ref.~\cite{Alioli:2015toa}.

For the parton shower, we use \pythia8.215~\cite{Sjostrand:2014zea} with different tunes. The default tune shown in the following is the \cms tune MonashStar~\cite{Khachatryan:2015pea} (tune 18 in \pythiaEight), also known as  CUETP8M1-NNPDF2.3LO, an underlying-event tune based on the Monash 2013 tune~\cite{Skands:2014pea}.  

In the following plots, we show in red (blue) the results of \geneva{}+\pythiaEight with (without) the MPI model turned on. 
For comparison, we also show the results of standalone \pythia using the same default tune as a green histogram.
The statistical uncertainties associated with Monte Carlo integration are not shown, since they are negligible for most of the observables presented below.
The perturbative uncertainty associated with scale variations is shown as a band of the same color as the corresponding \geneva central prediction. These should be interpreted as the theory uncertainty on the perturbative part of our calculation, which only includes the effects of the primary interaction, and it is therefore reported for illustrative purposes. For many observables we will show, the dominant part of the actual theory uncertainty will be given by secondary interactions and nonperturbative physics.

To estimate these effects, we also provide results for the \atlas UE Tune AU2-CT10~\cite{ATL-PHYS-PUB-2012-003} (tune 11 in \pythiaEight) and the \pythia8.215 default tune Monash 2013 (tune 14), which were chosen to represent the typical range of tune variations. We also report the result for \atlas Tune AZ~\cite{Aad:2014xaa} (tune 17 in \pythiaEight). At variance with the other tunes, the \atlas Tune AZ started from tune 4C and uses data for the $Z$-boson transverse momentum,  without explicitly including recent UE-sensitive measurements from the LHC. We therefore expect its agreement with data to be less optimal for UE-sensitive measurements, but we report it as an indication of the uncertainties associated with the choice of tune.  In the following plots, the central values for these other tunes are shown as dashed lines of similar color as the corresponding \geneva or \pythia prediction for tune MonashStar.

Finally, in order to limit the contamination from  different physical effects and to keep the event record as simple as possible both  \geneva and \pythiaEight predictions do not include any QED or EW showering.

\subsection{\atlas}
The underlying event observables used in the comparison with \atlas data are based on the charged particle tracks in each event. The transverse momentum of the muon pair is used as a measure for the hardness of the event. Observables are shown as differential distributions in a certain \ZpT range, or their mean value is plotted as a function of \ZpT in profile histograms. Each observable is divided in three different angular regions defined with respect to the $Z$-boson direction. The toward region subtends an azimuthal cone of angle $2\pi/3$ around the $Z$-boson direction. The away region is the cone of azimuthal size $2\pi/3$ directly opposite the toward region, and the transverse region covers the remainder of the solid angle.
The transverse region is further separated on an event-by-event basis by measuring the sum of the charged-particle transverse momenta in each one of the two opposite contributing regions. The region with more activity is labelled ``trans-max'' while the other is labelled ``trans-min''. The difference between them for a given observable is usually referred to as the ``trans-diff'' region~\cite{Marchesini:1988hj,Pumplin:1997ix}, although it does not actually represent a region. This additional separation is helpful because while soft UE effects are expected to contribute equally to the ``trans-min'' and ``trans-max'' regions, further primary perturbative radiation should contribute more to ``trans-max''. Therefore, in the ``trans-diff'' contributions to observables the UE effects are expected to largely cancel enhancing the sensitivity to the primary radiation component.

The observables we consider are the number of charged particles per unit $\eta$--$\phi$, \Nchgdetadphi, the scalar \pt sum of stable charged particles per unit $\eta$--$\phi$, \pTsumdetadphi, and the average \pt of charged particles, \ptmean.

\begin{figure*}[!ht]
\begin{center}
\includegraphics[width=0.5\textwidth]{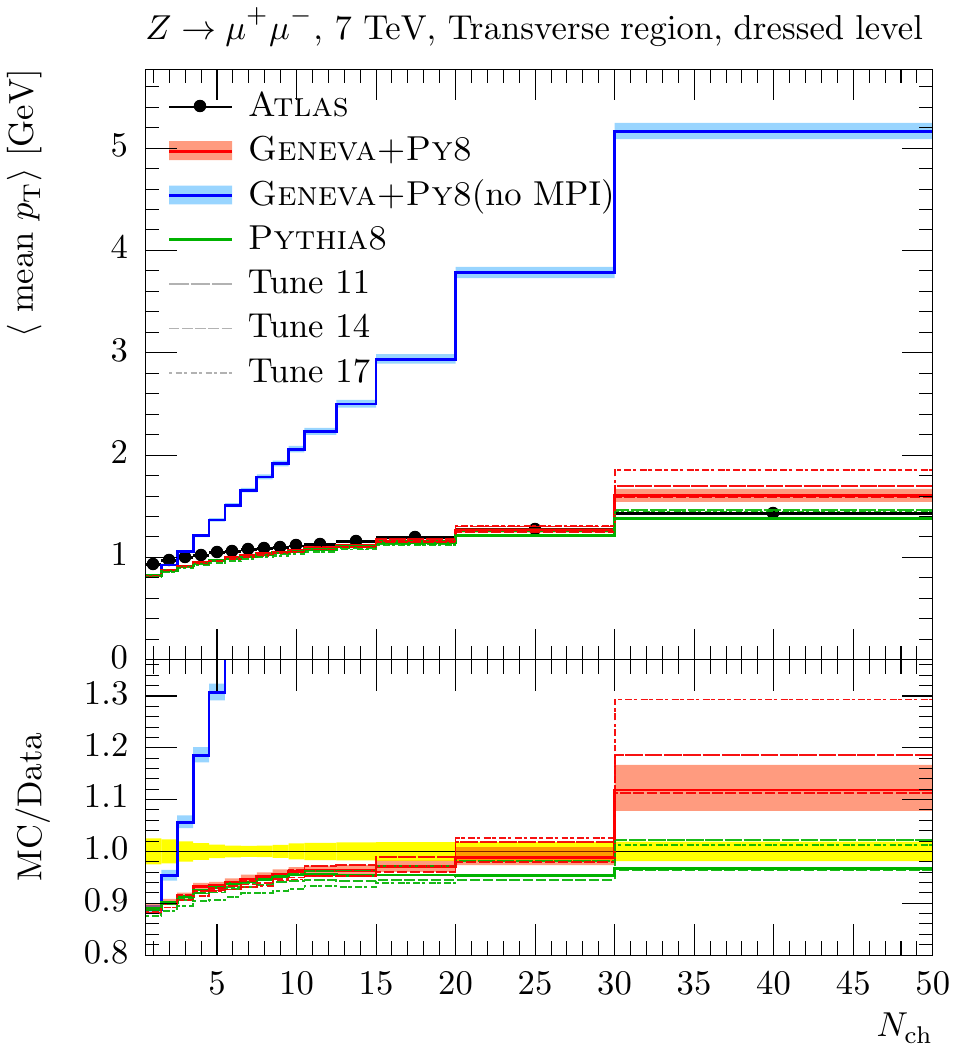}%
\hfill%
\includegraphics[width=0.5\textwidth]{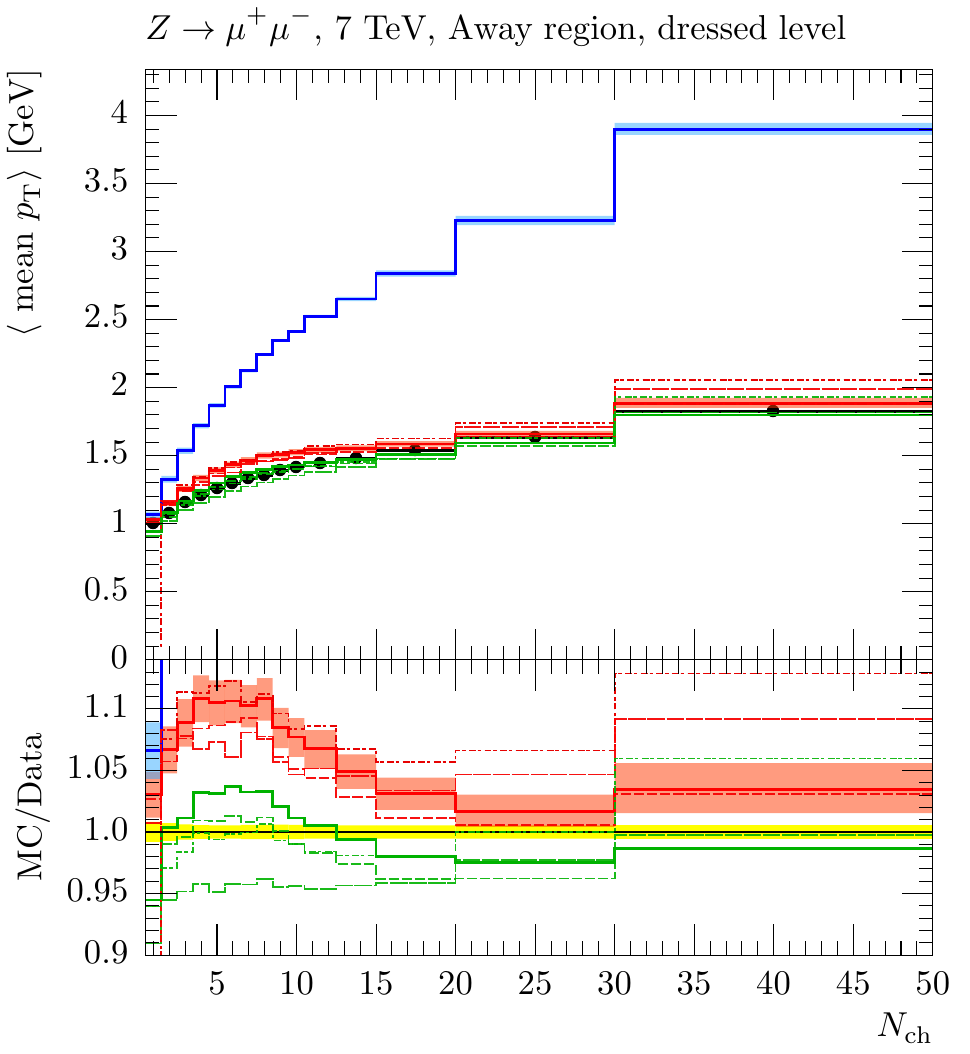}%
\caption{\label{fig:ATLAS_MeanPT}Mean charged particle \pt as a function of multiplicity, in the transverse region (left panel) and away region (right panel).}
\end{center}
\end{figure*}

Figure \ref{fig:ATLAS_MeanPT} shows the mean \pt per charged particle as a function of charged particle multiplicity, both in the transverse and the away region. \geneva without MPI effects has a much larger \ptmean for events with many charged particles than observed and is clearly inadequate to describe the data. This is because MPI adds many soft charged tracks to collision events, which has the effect of lowering the mean \pt per charged particle. However, with MPI turned on, \geneva describes the data quite well, and its agreement is roughly at the level of \pythia standalone. 

\begin{figure*}[!ht]
\begin{center}
\includegraphics[width=0.5\textwidth]{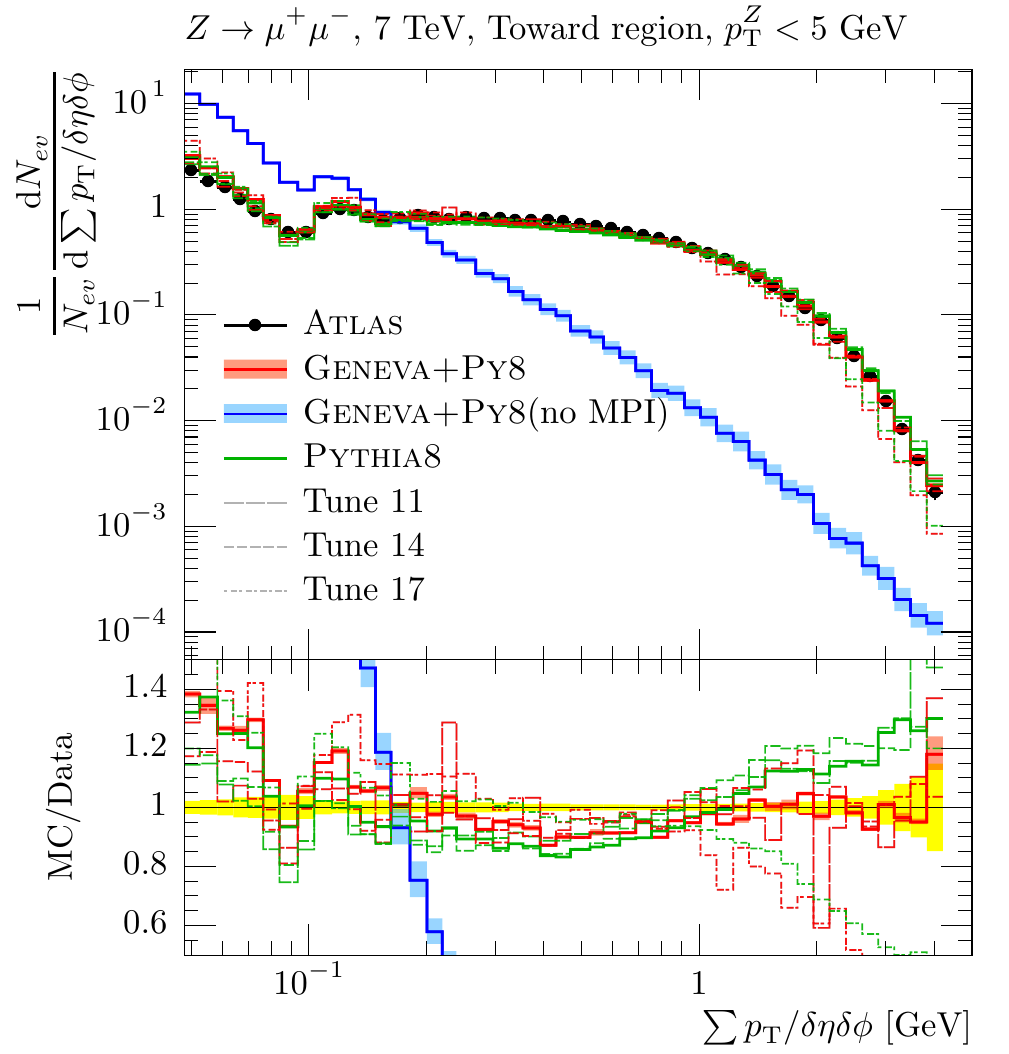}%
\hfill%
\includegraphics[width=0.5\textwidth]{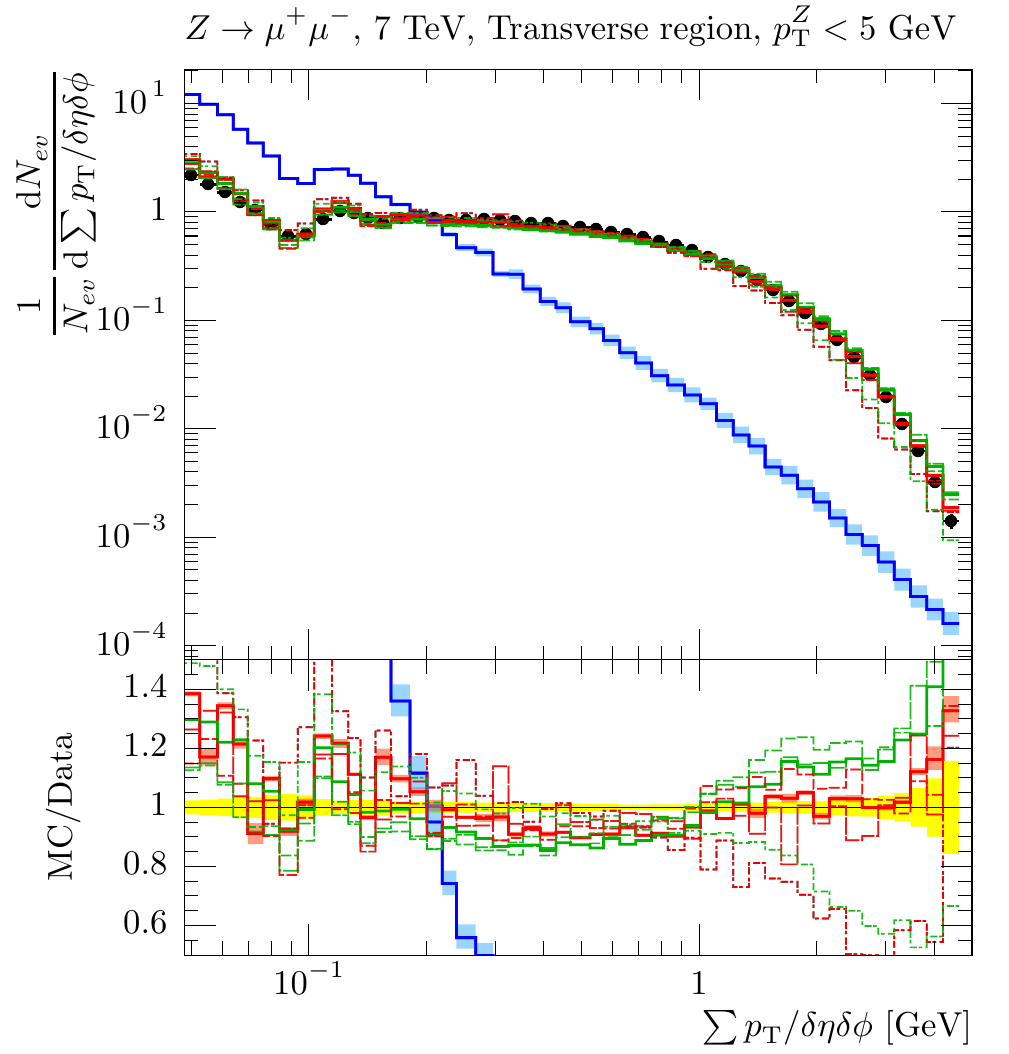}%
\caption{\label{fig:ATLAS_dSumPT_lowZpT}The differential \pTsumdetadphi distribution at low \ZpT $ < 5\GeV$, in the toward region (left panel) and transverse region (right panel).}
\end{center}
\end{figure*}

Figure \ref{fig:ATLAS_dSumPT_lowZpT} shows the differential \dpTsumdetadphi distribution for $\ZpT < 5 \GeV$. At such low values of \ZpT one expects the extra perturbative effects included in \geneva to matter less than MPI and nonperturbative effects. As expected, \geneva without MPI clearly fails to capture the shape of the distribution, and it undershoots it for $\sum p_T > 200$ MeV. Once MPI is turned on, \geneva agrees well with standalone \pythia both in the toward and transverse region.
Among the various tunes, \atlas tune AZ (tune 17) clearly performs worse, which reflects the lack of updated UE-sensitive information in its input.
The other tunes agree very well with the data, which clearly supports that the procedure for interfacing \geneva to \pythia does not spoil the shower accuracy of \pythia. 

\begin{figure*}[!ht]
\begin{center}
\includegraphics[width=0.5\textwidth]{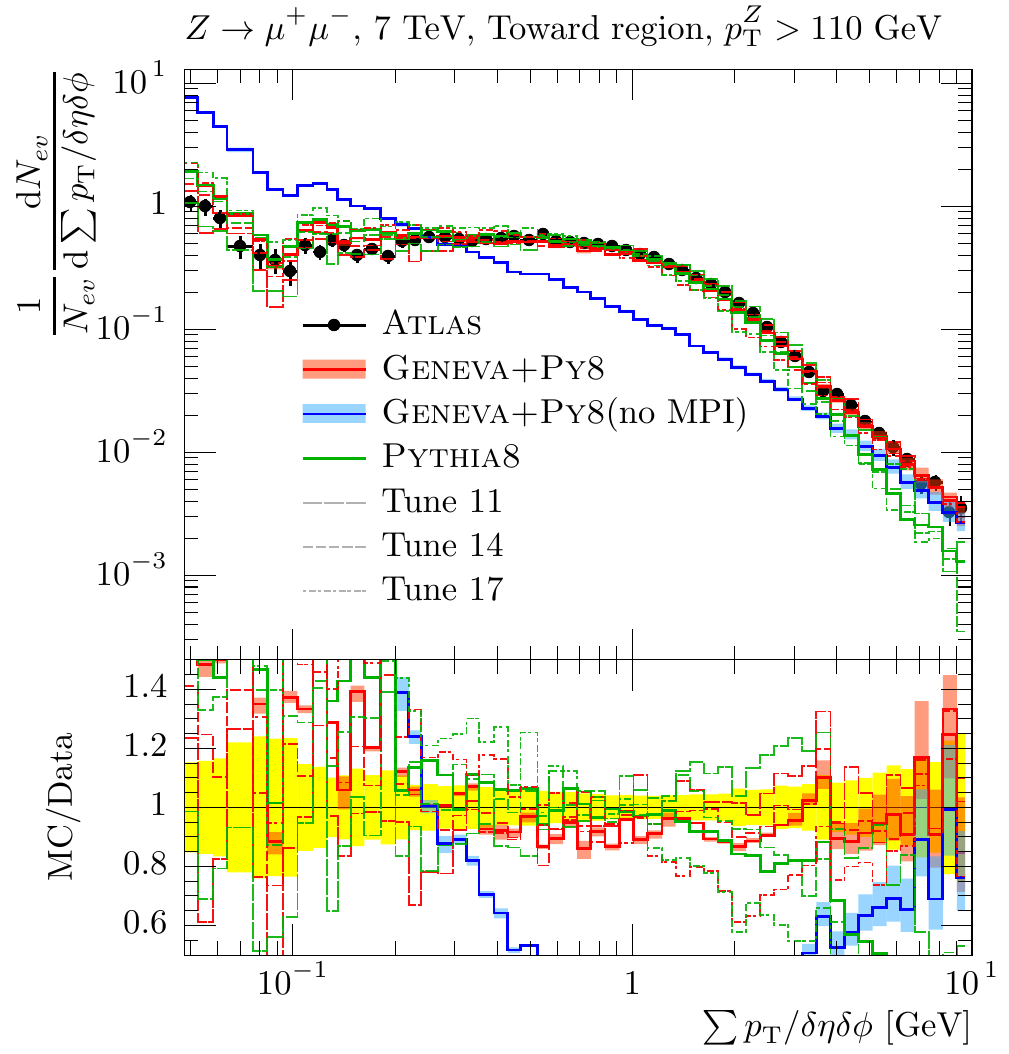}%
\hfill%
\includegraphics[width=0.5\textwidth]{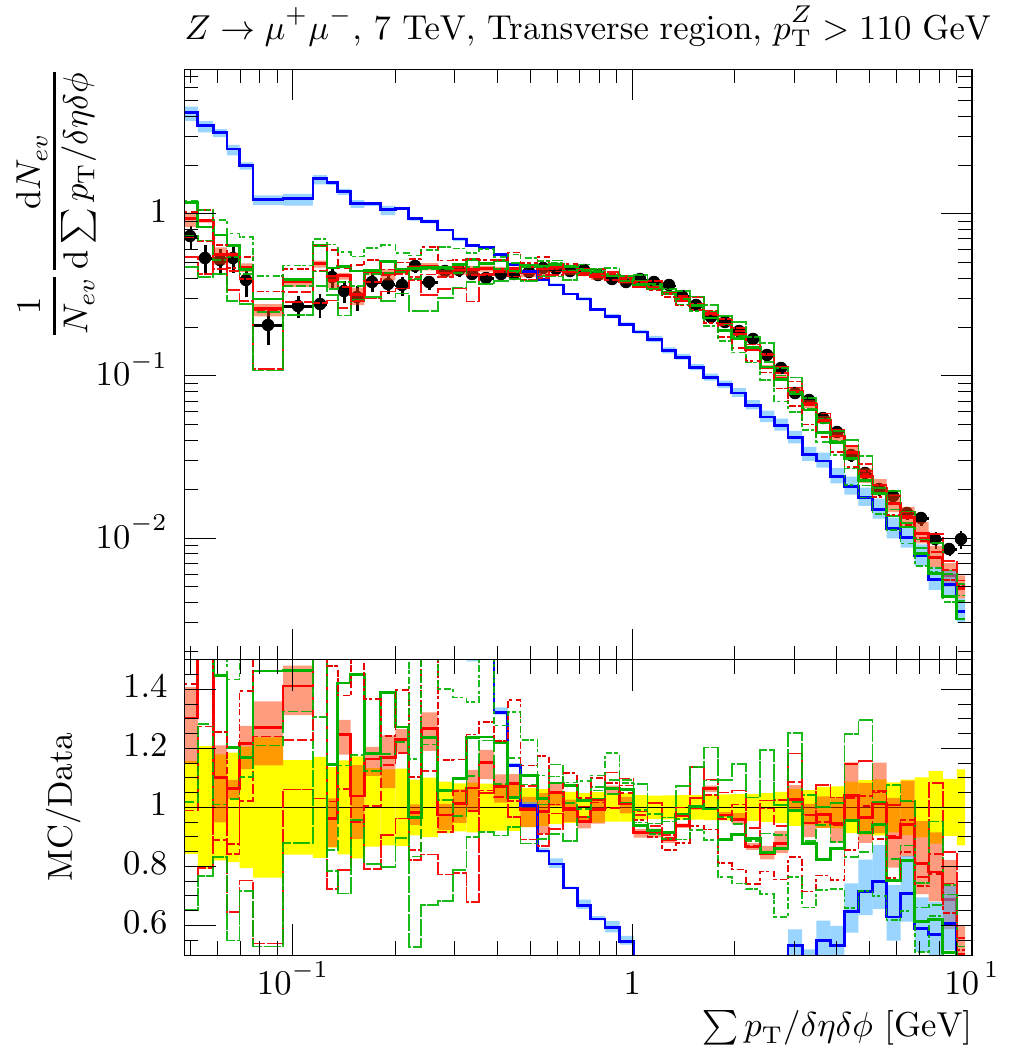}%
\caption{\label{fig:ATLAS_dSumPT_highZpT}The differential \pTsumdetadphi distribution at high \ZpT $ > 110 \GeV$, in the toward region (left panel) and transverse region (right panel).}
\end{center}
\end{figure*}

For $\ZpT > 110 \GeV$,  as shown in \fig{ATLAS_dSumPT_highZpT}, \geneva and \pythia both agree well with each other (within the larger statistical fluctuations that appear in presence of such a hard cut). They also agree well with the data in the transverse region, while in the toward region \geneva gives a better description of the data at high values of \dpTsumdetadphi. The better agreement of \geneva could be due to the more accurate perturbative information included, however a more detailed understanding of the interplay between the perturbative and nonperturbative physics would be required to make a definite statement.

\begin{figure*}[!ht]
\begin{center}
\includegraphics[width=0.5\textwidth]{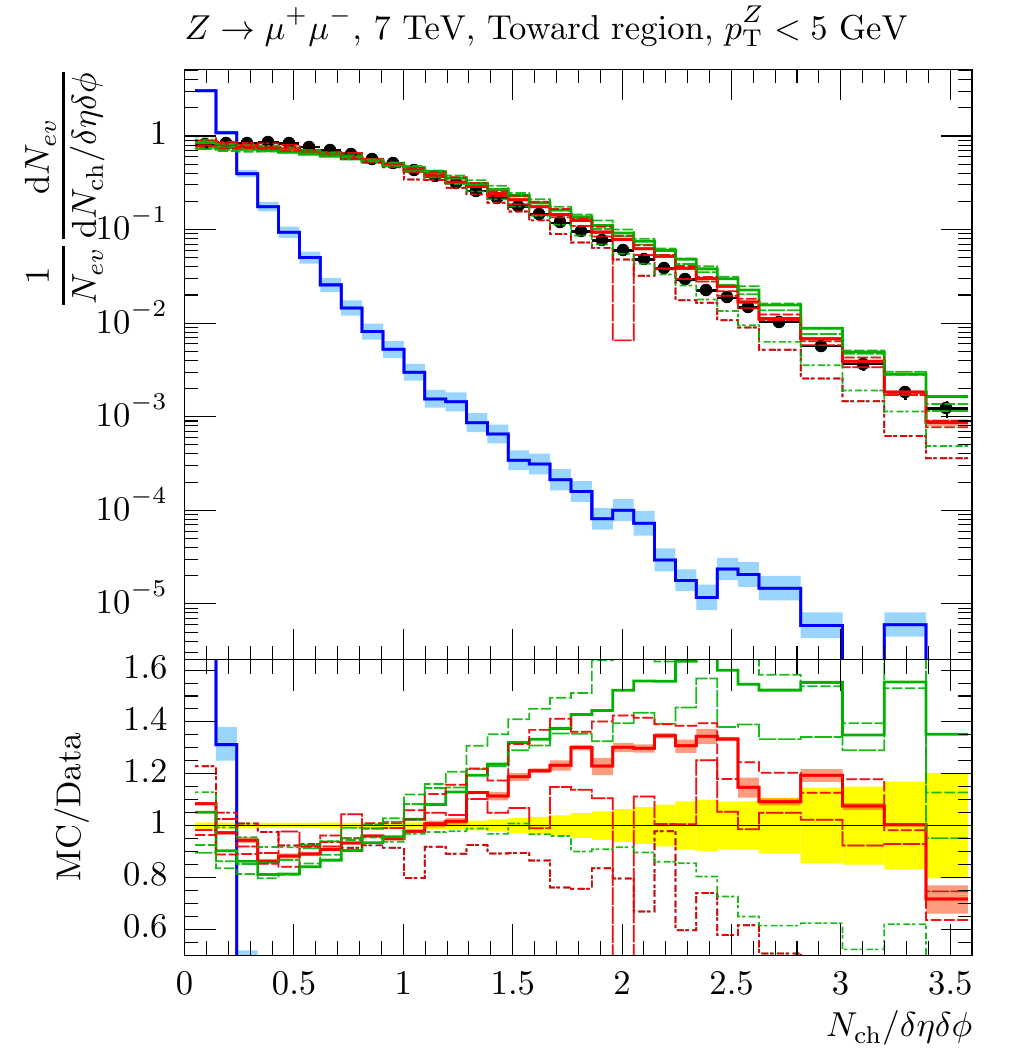}%
\hfill%
\includegraphics[width=0.5\textwidth]{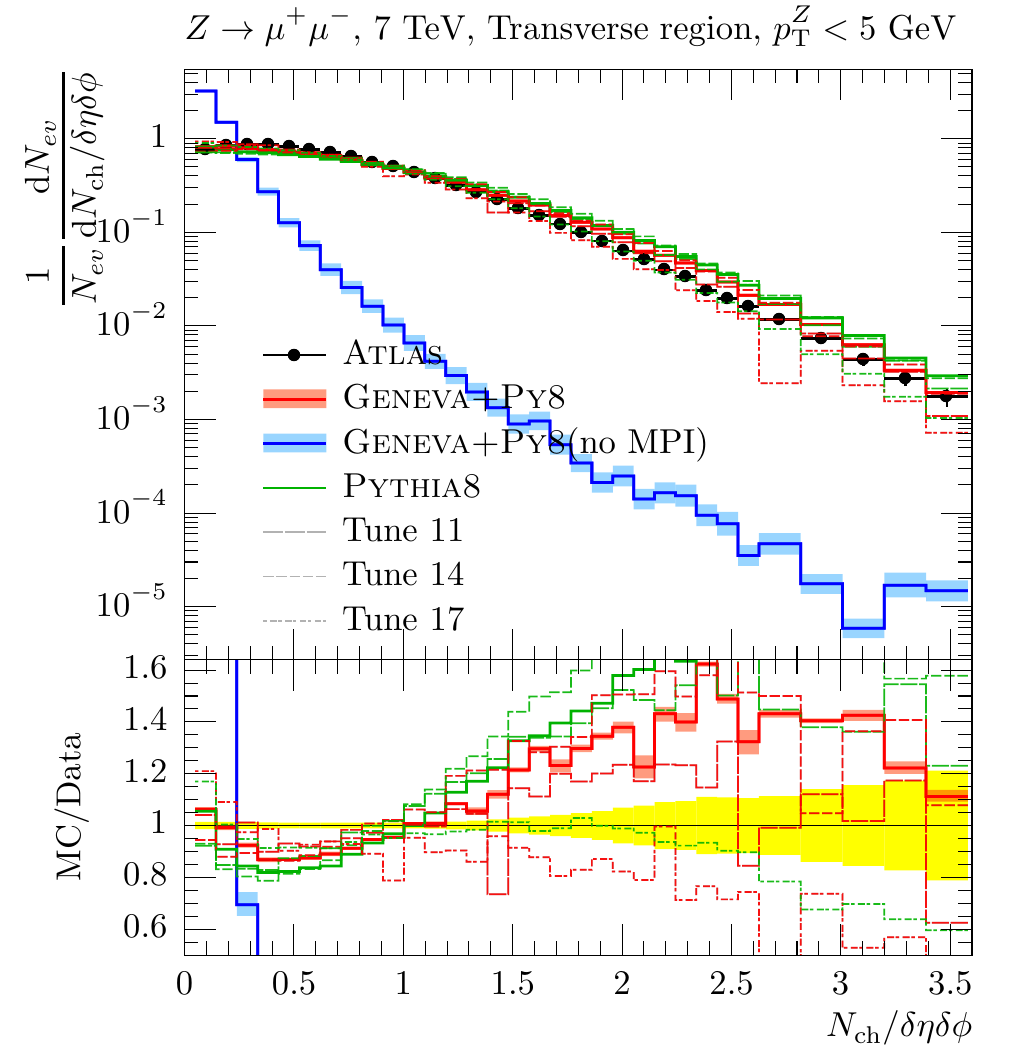}%
\caption{\label{fig:ATLAS_dNchg_lowZpT}The differential \Nchgdetadphi distribution at low \ZpT $ < 5\GeV$, in the toward region (left panel) and transverse region (right panel).}
\end{center}
\end{figure*}

\begin{figure*}[!ht]
\begin{center}
\includegraphics[width=0.5\textwidth]{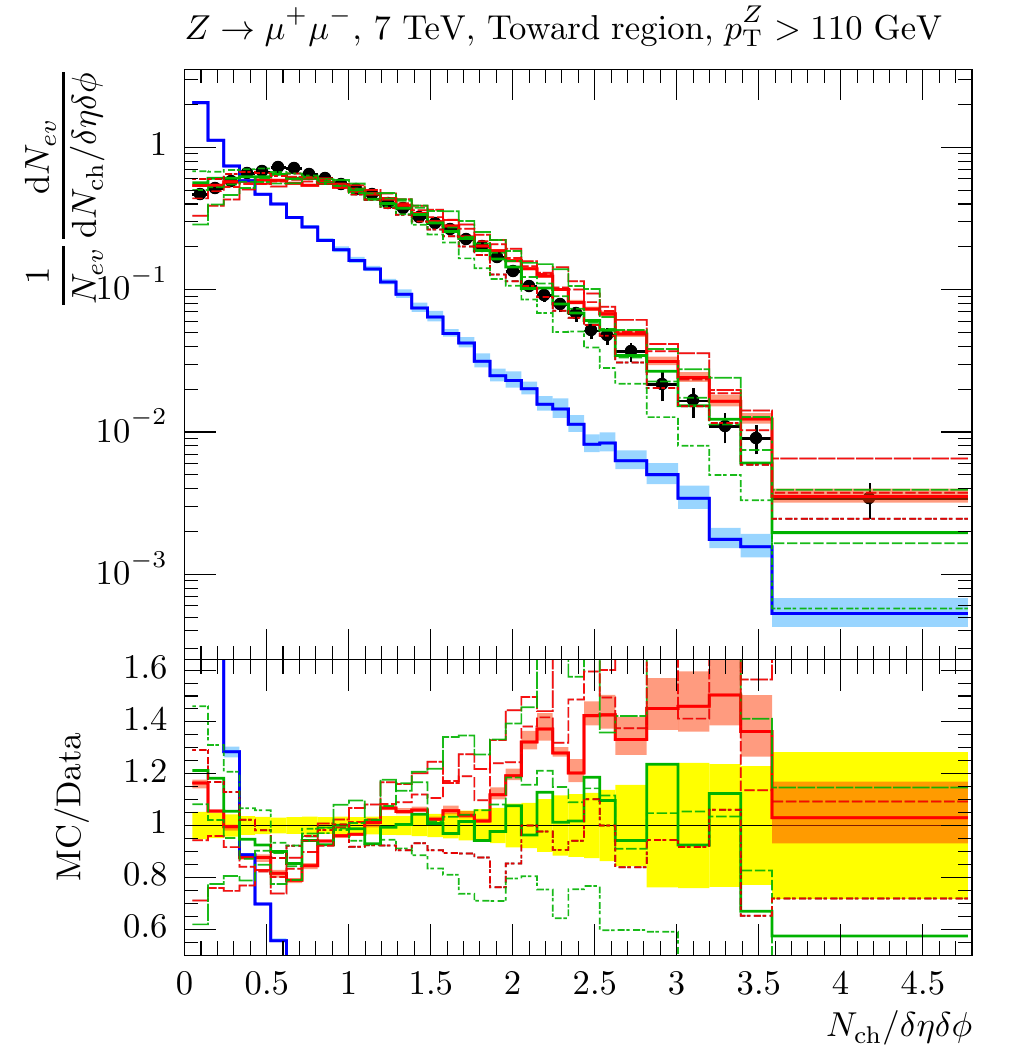}%
\hfill%
\includegraphics[width=0.5\textwidth]{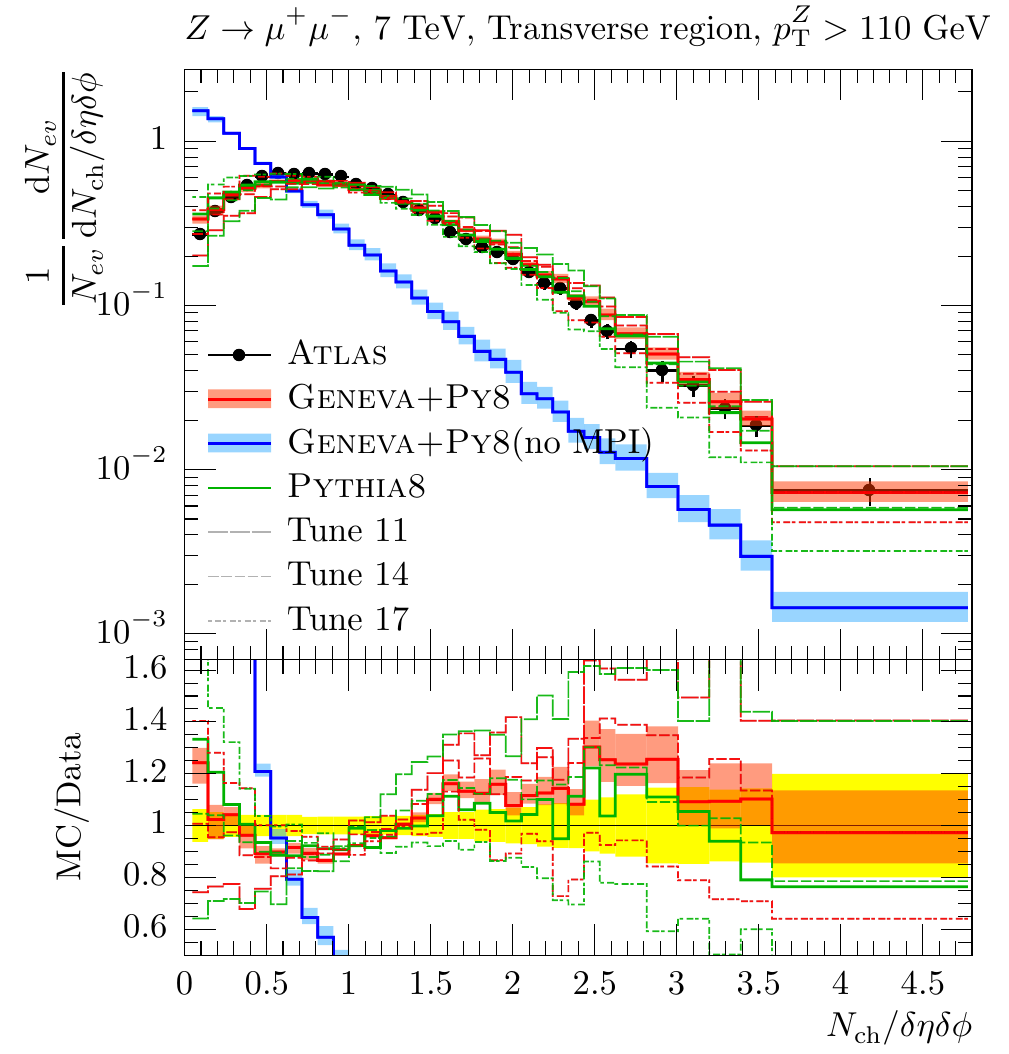}%
\caption{\label{fig:ATLAS_dNchg_highZpT}The differential \Nchgdetadphi distribution at high \ZpT $ > 110\GeV$, in the toward region (left panel) and transverse region (right panel).}
\end{center}
\end{figure*}

Similar results are obtained for the \dNchgdetadphi distributions. For $\ZpT < 5 \GeV$ shown in \fig{ATLAS_dNchg_lowZpT}, \geneva without MPI is clearly not giving enough charged tracks, while both \geneva with MPI and \pythia standalone give roughly the same predictions (both disagree with the data somewhat). For $\ZpT > 110 \GeV$,  as shown in \fig{ATLAS_dNchg_highZpT}, differences between \geneva and standalone \pythia develop again. For the central tune we have chosen, \pythia agrees slightly better with the data, however, the uncertainties from the \pythia tunes are large. Again, \atlas tune AZ (tune 17) seems to be in worse agreement with data.

\begin{figure*}[!ht]
\begin{center}
\includegraphics[width=0.48\textwidth]{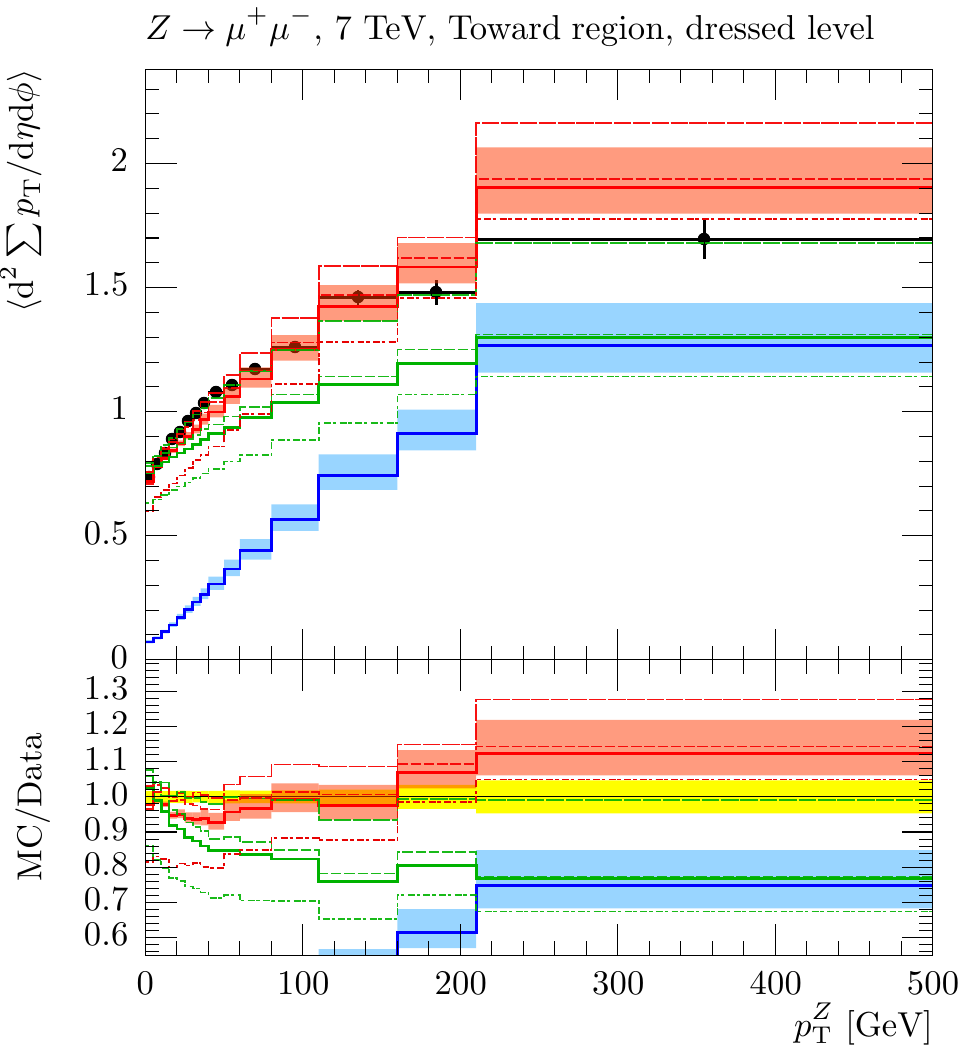}%
\hfill%
\includegraphics[width=0.48\textwidth]{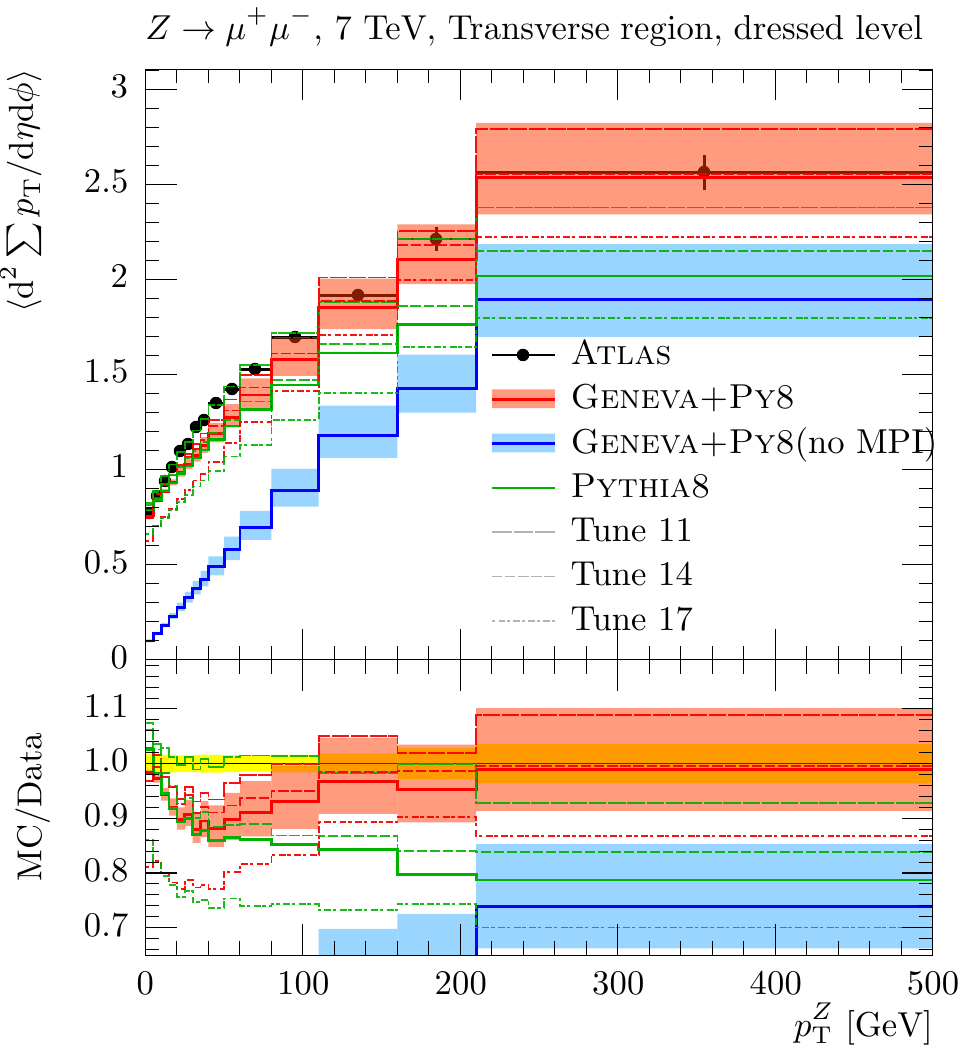}%
\caption{\label{fig:ATLAS_SumPT}The charged particle scalar $\sum p_T$ density average values, as a function of $Z$-boson transverse momentum $p_T^Z$, in the toward region (left panel) and transverse region (right panel).}

\end{center}
\end{figure*}

\begin{figure*}[!ht]
\begin{center}
\includegraphics[width=0.48\textwidth]{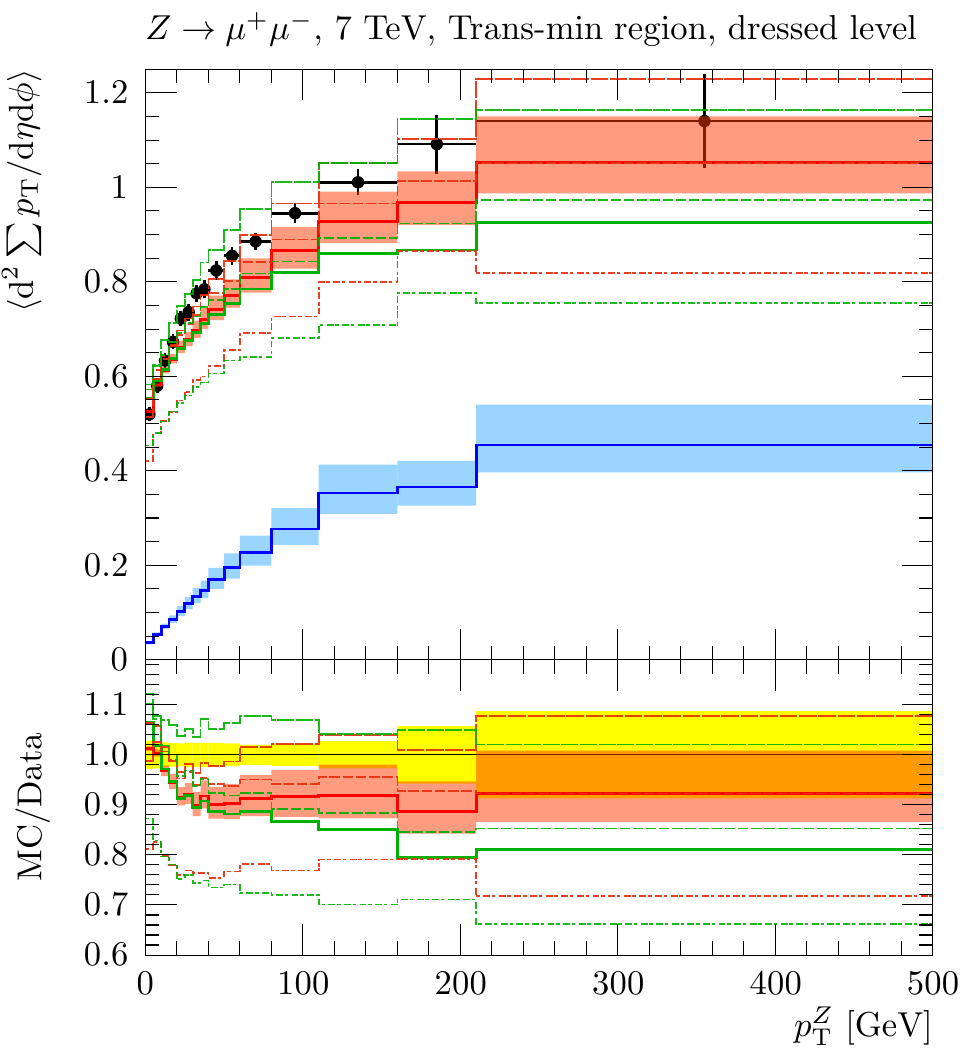}%
\hfill%
\includegraphics[width=0.48\textwidth]{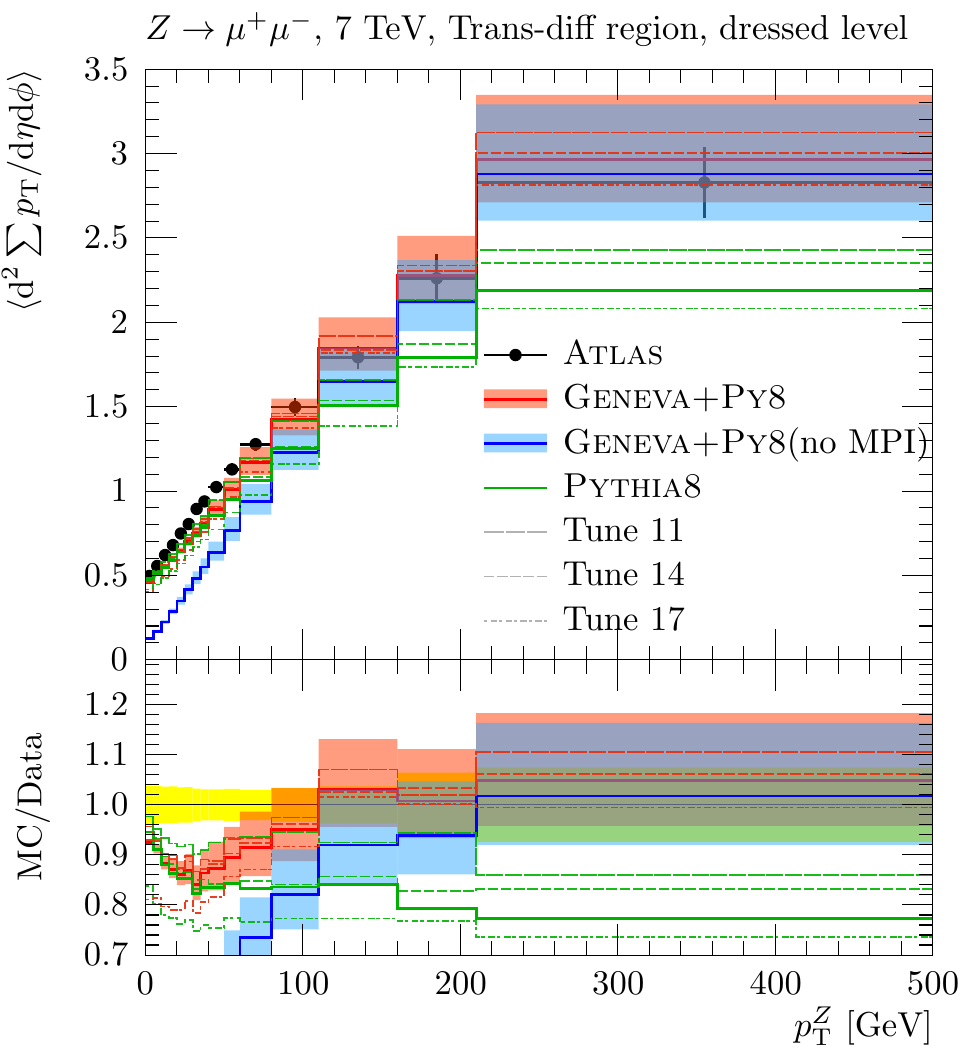}%
\caption{\label{fig:ATLAS_SumPTdiff}The charged particle scalar $\sum p_T$ density average values, as a function of $Z$-boson transverse momentum $p_T^Z$, in the trans-min region (left panel) and trans-diff region (right panel).}

\end{center}
\end{figure*}

Averaging each observable allows one to look at the description of the
data across the whole \ZpT spectrum. Figure \ref{fig:ATLAS_SumPT}
shows the average of \dpTsumdetadphi as a function of $\ZpT$.  As
before, \geneva without MPI does not describe the data at all. \pythia
on its own falls below the data as $\ZpT$ increases, whereas \geneva
stays closer to the data even for very high $\left( > 100 \GeV\right)$
bins of \ZpT in both regions.  This is also evident in Figure
\ref{fig:ATLAS_SumPTdiff}, which shows the same obervable in the
``trans-min'' and ``trans-diff'' regions. While the tuning of \pythia manages to bring
it into fair agreement with data in the more UE-sensitive ``trans-min'' region, the lack of an higher-order description of primary radiation results in \pythia undershooting the measurements at high $\ZpT$ in the less UE-sensitive ``trans-diff'' region. The predictions of \geneva are instead in good agreement with the data in both regions.

\begin{figure*}[!ht]
\begin{center}
\includegraphics[width=0.48\textwidth]{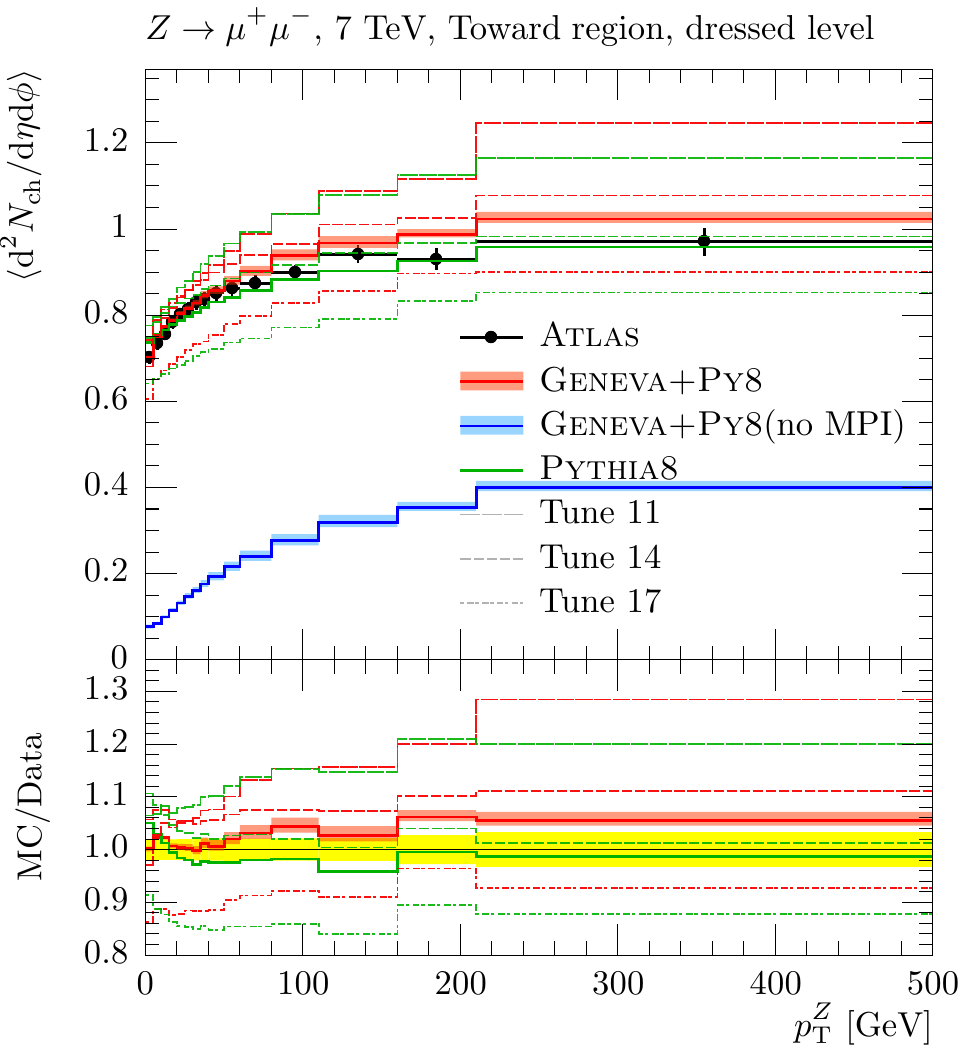}%
\hfill%
\includegraphics[width=0.48\textwidth]{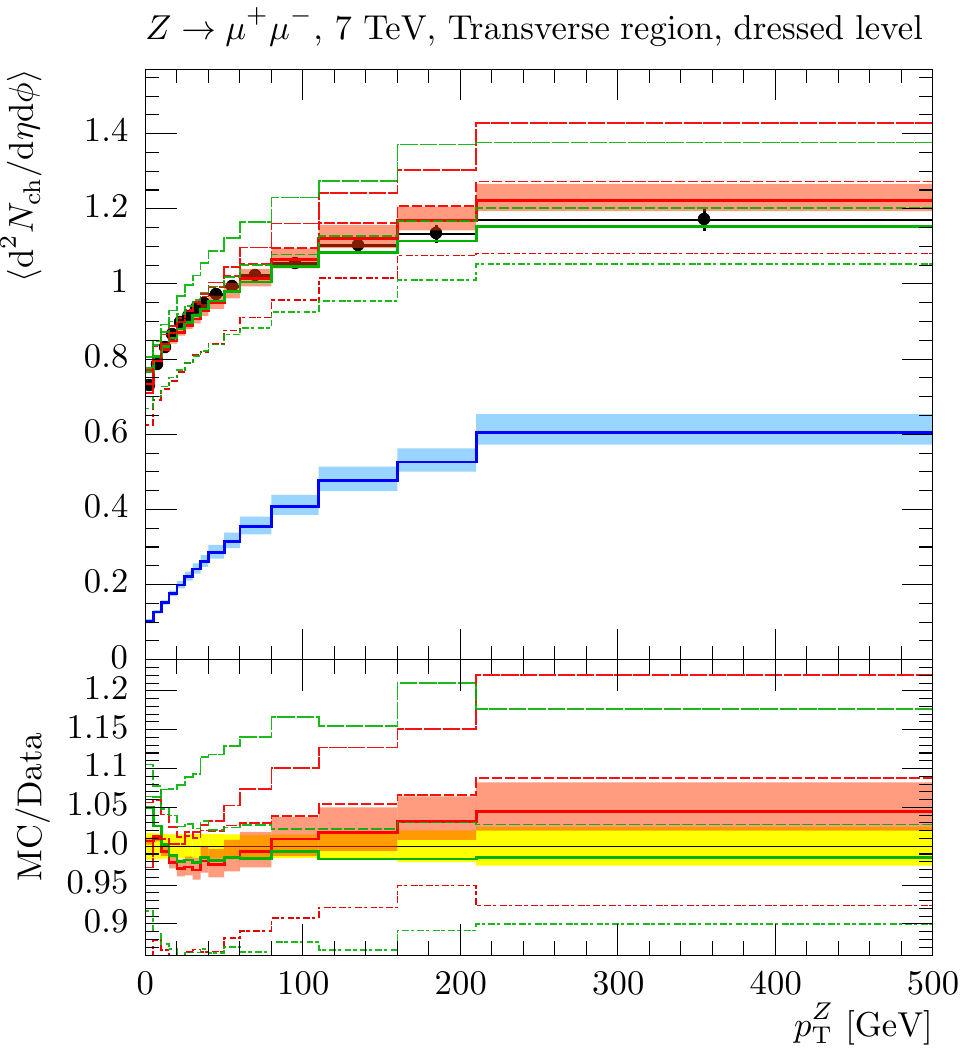}%
\caption{\label{fig:ATLAS_Nchg}
The number of charged particle tracks, as a function of $Z$-boson transverse momentum $p_T^Z$, in the toward region (left panel) and transverse region (right panel).}
\end{center}
\end{figure*}

\begin{figure*}[!ht]
\begin{center}
\includegraphics[width=0.48\textwidth]{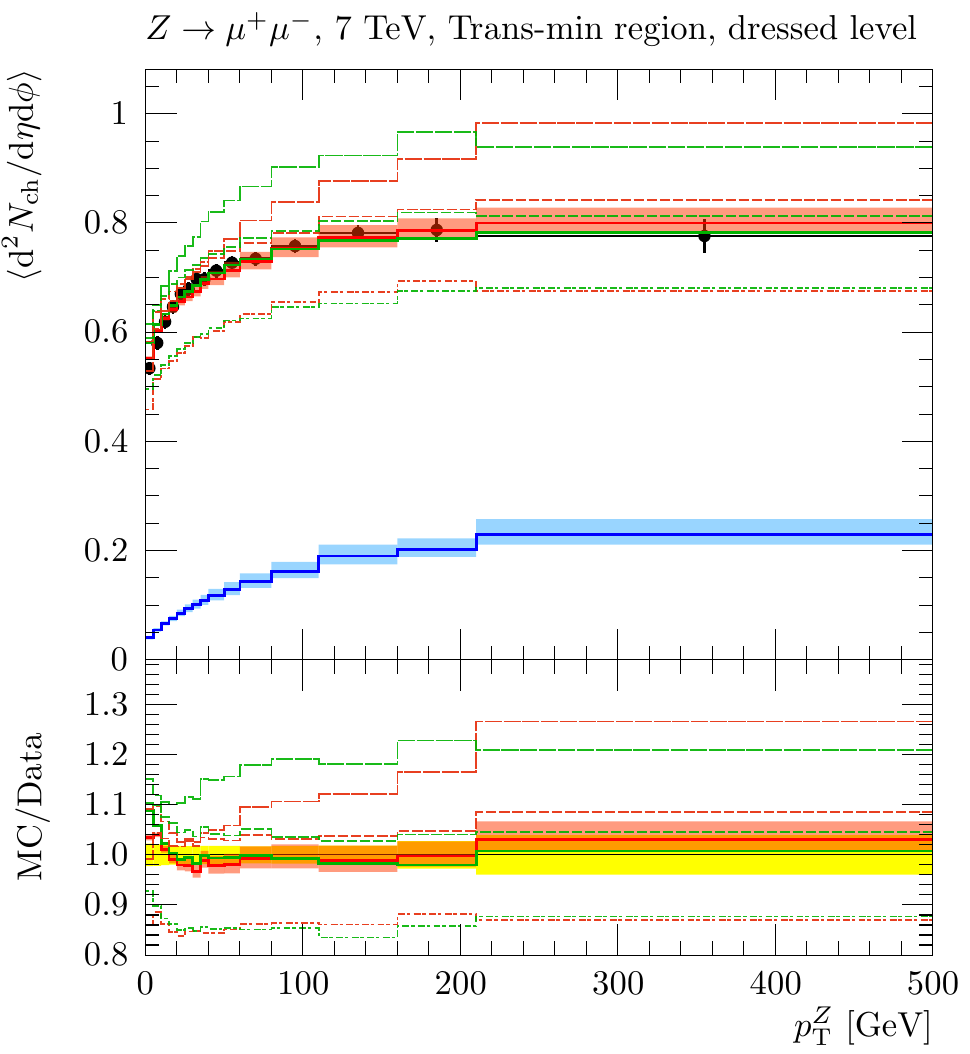}%
\hfill%
\includegraphics[width=0.48\textwidth]{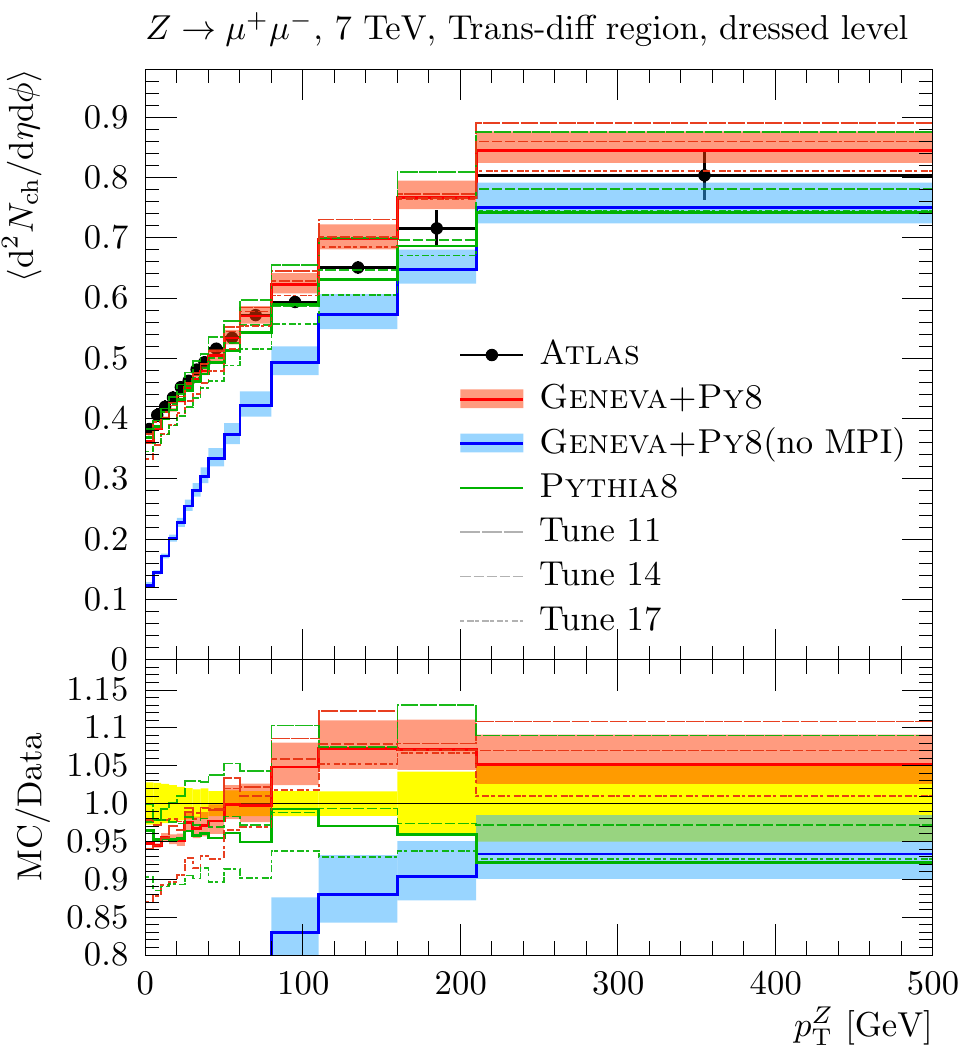}%
\caption{\label{fig:ATLAS_Nchgdiff}
The number of charged particle tracks, as a function of $Z$-boson transverse momentum $p_T^Z$, in the trans-min region (left panel) and trans-diff region (right panel).}
\end{center}
\end{figure*}

The \Nchgdetadphi distribution as a function of \ZpT in figures \ref{fig:ATLAS_Nchg} and  \ref{fig:ATLAS_Nchgdiff} follow a similar pattern, but the agreement of \pythiaEight with the data is overall much better than for the average \dpTsumdetadphi distribution, and the difference between \geneva and \pythia is less pronounced. Both are in reasonable agreement with the data.
In the ``trans-diff'' region, both \pythia standalone and \geneva without MPI approach the data at high $\ZpT$, while \geneva with MPI now overshoots the data. This is a clear indication that some primary effects have been tuned into \pythia's MPI model.

\subsection{\cms}
\begin{figure*}[!ht]
\begin{center}
\includegraphics[width=0.48\textwidth]{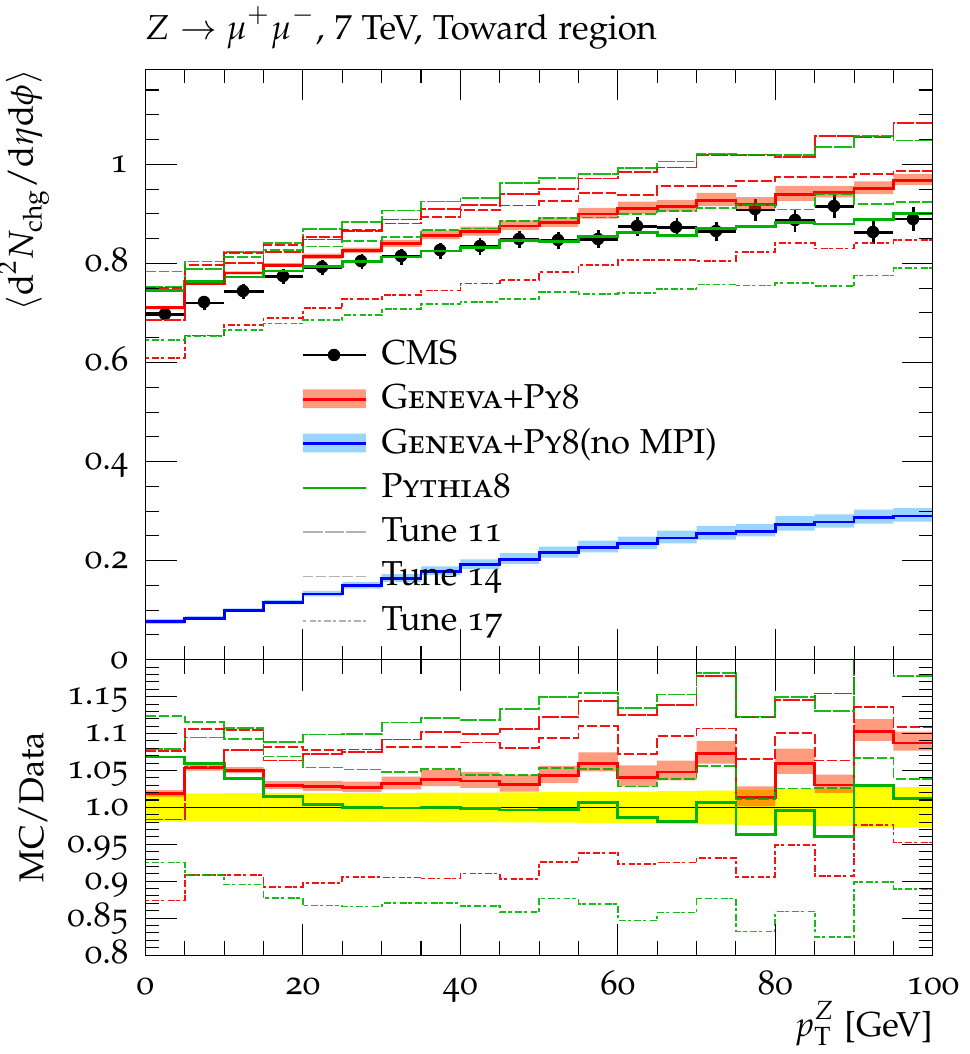}%
\hfill%
\includegraphics[width=0.48\textwidth]{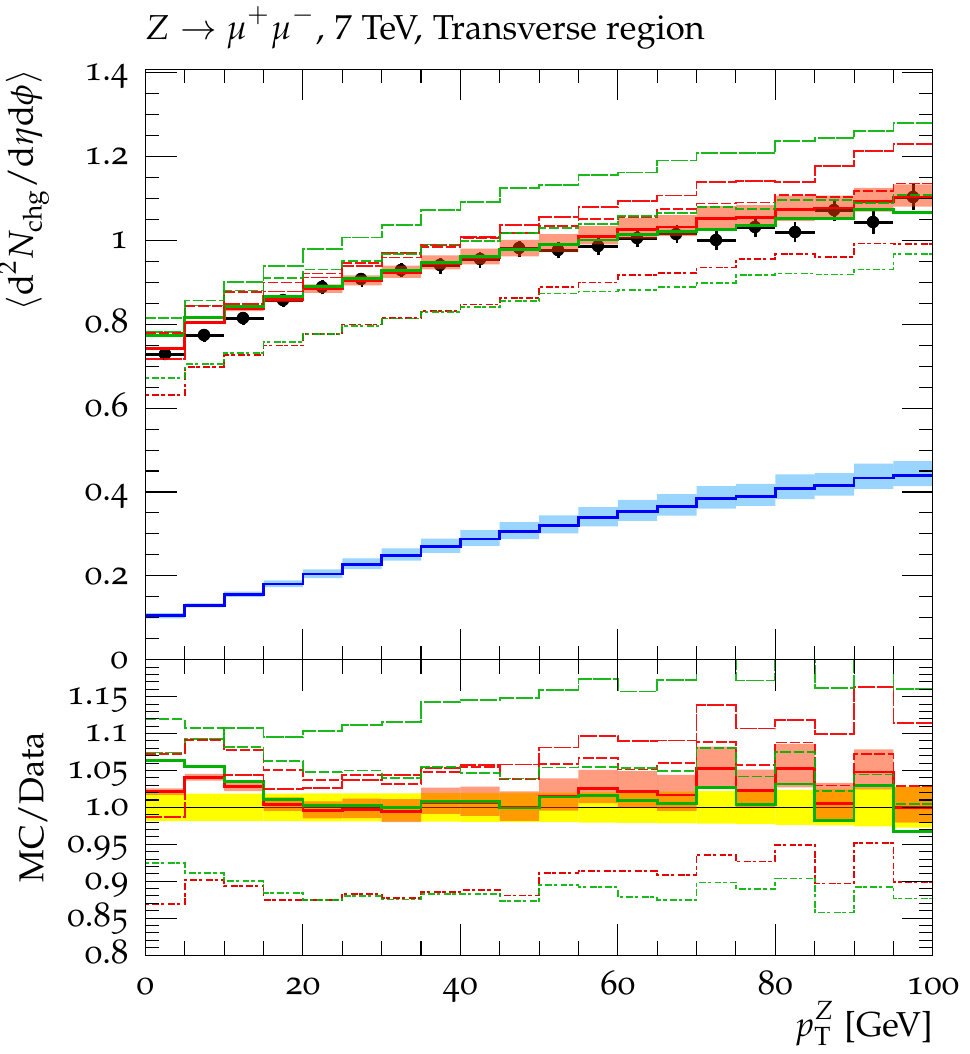}%
\hfill%
\includegraphics[width=0.48\textwidth]{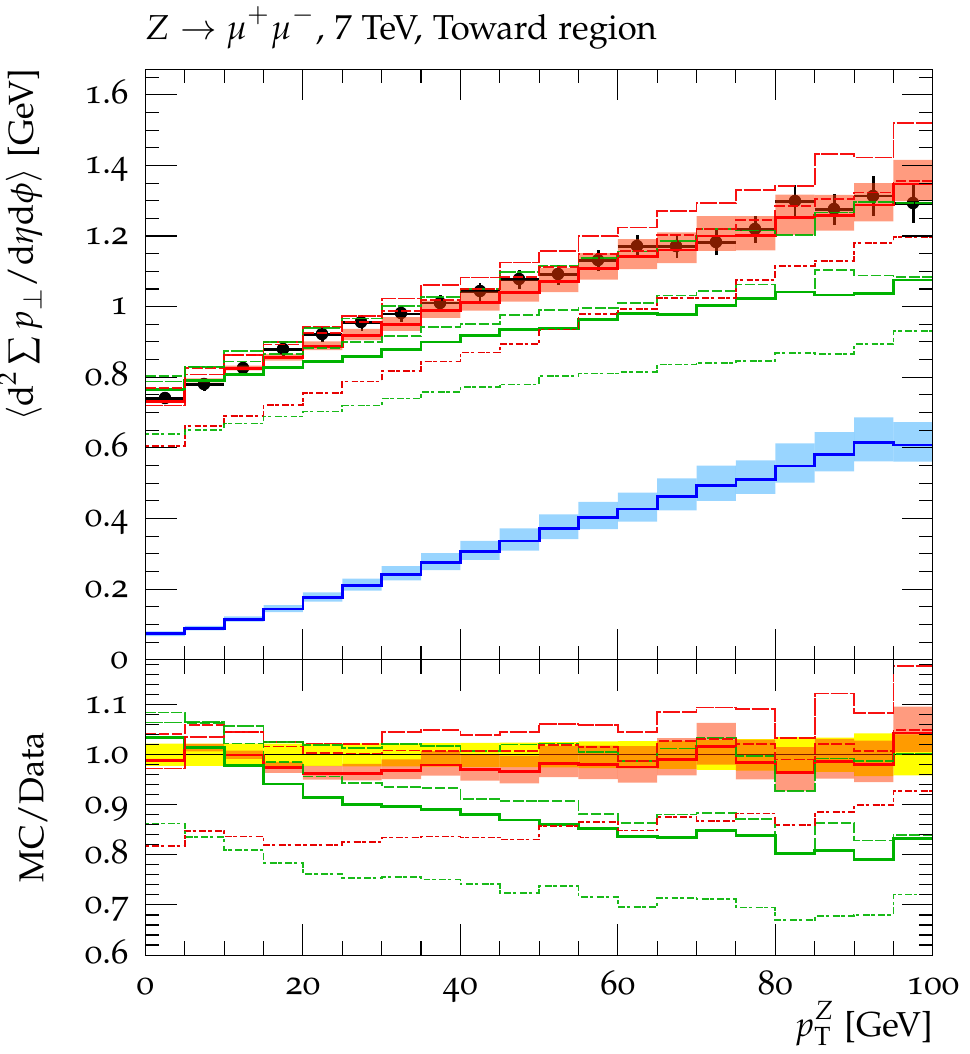}%
\hfill%
\includegraphics[width=0.48\textwidth]{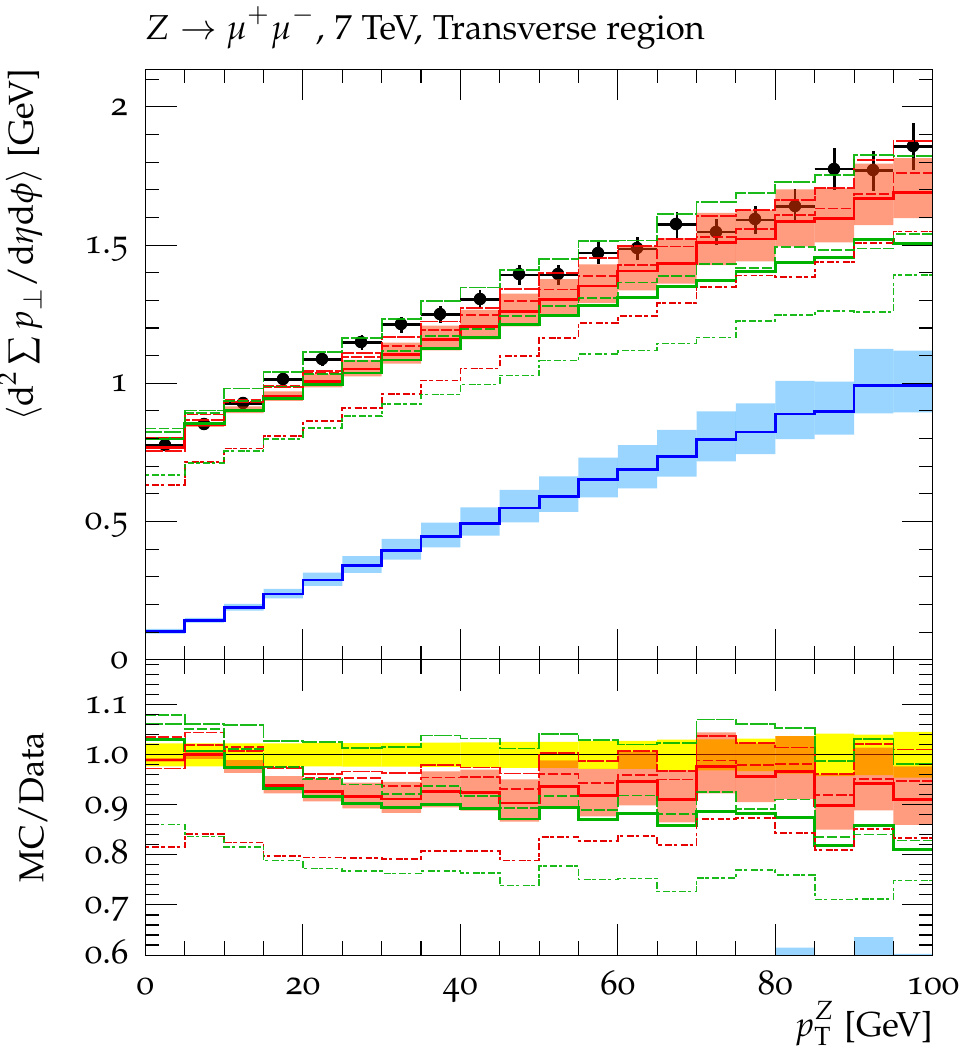}%
\caption{\label{fig:CMS_Nchg_SumPT}
The number of charged particle tracks (top) and summed charged particle transverse momentum (bottom) as a function of $Z$-boson transverse momentum $p_T^Z$ in the toward region (left panels) and transverse region (right panels).}
\end{center}
\end{figure*}

\begin{figure*}[!ht] 
\begin{center}
\includegraphics[width=0.48\textwidth]{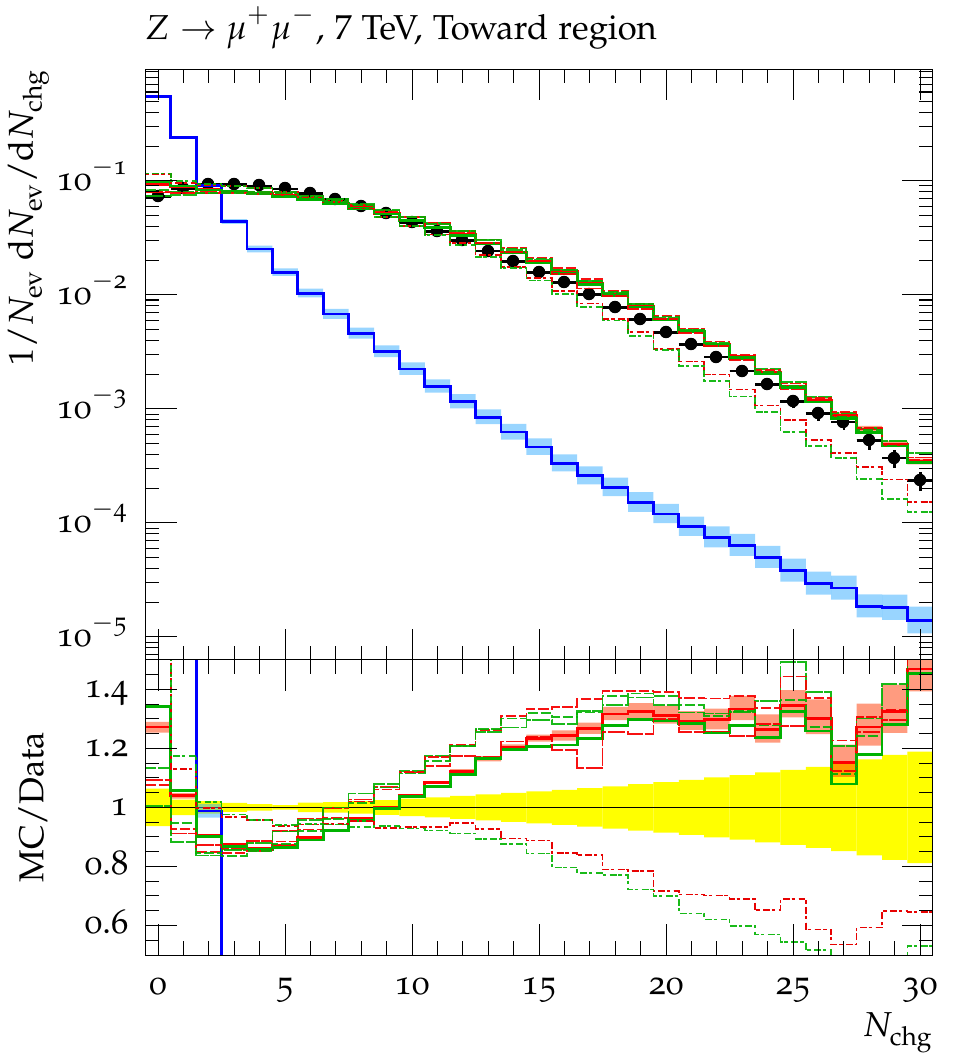}%
\hfill%
\includegraphics[width=0.48\textwidth]{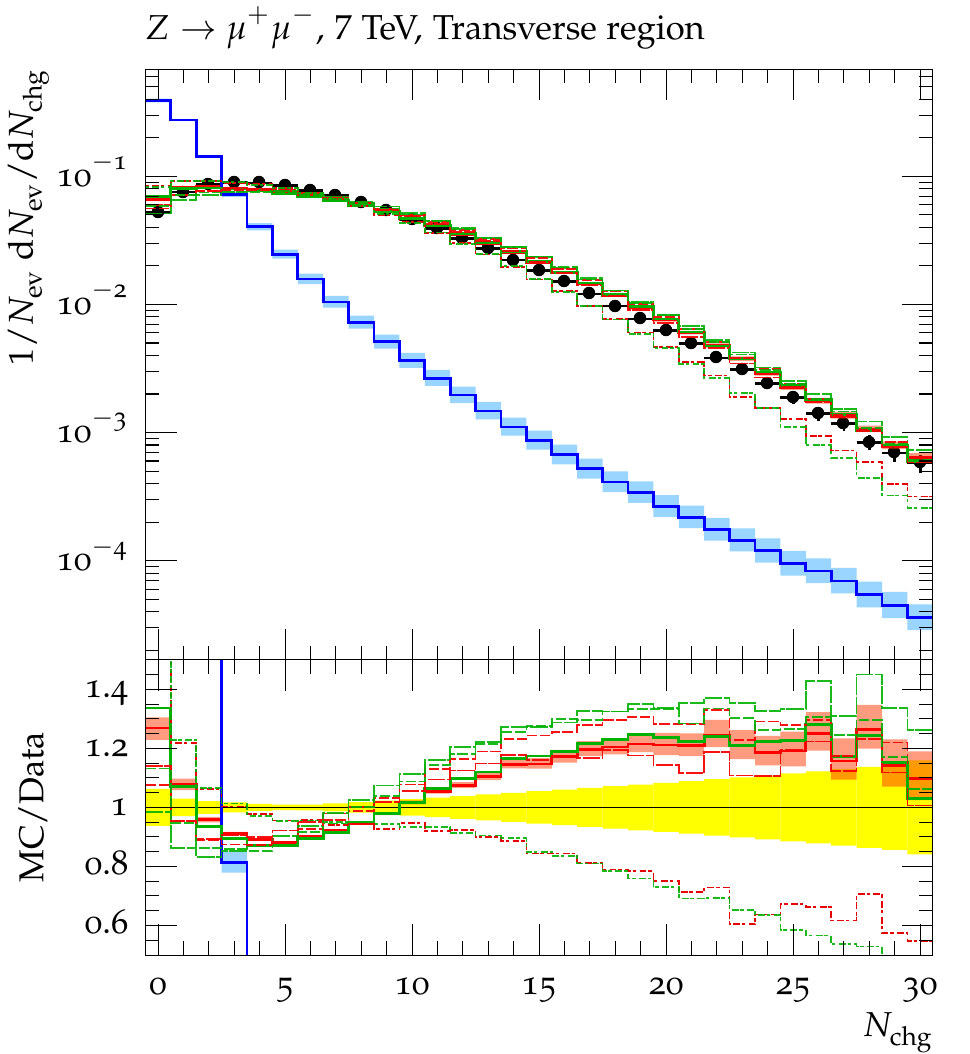}%
\hfill%
\includegraphics[width=0.48\textwidth]{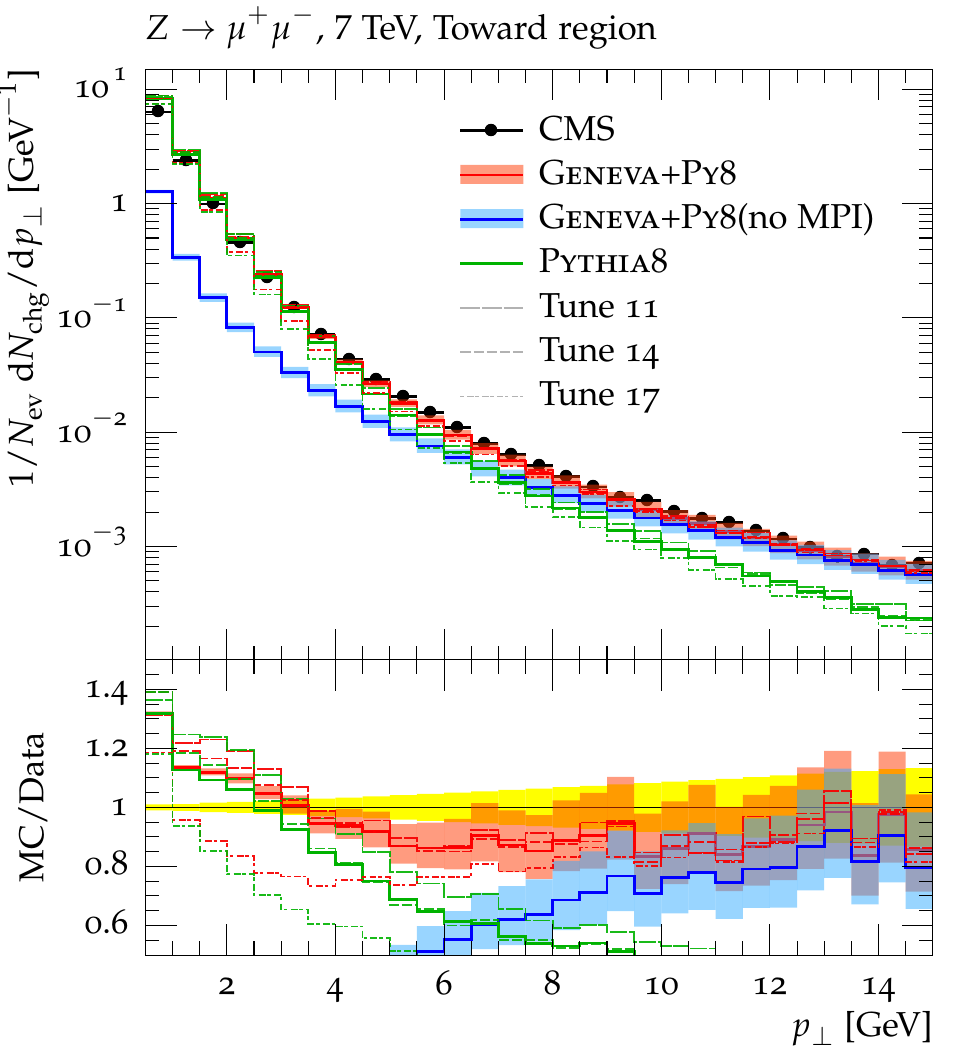}%
\hfill%
\includegraphics[width=0.48\textwidth]{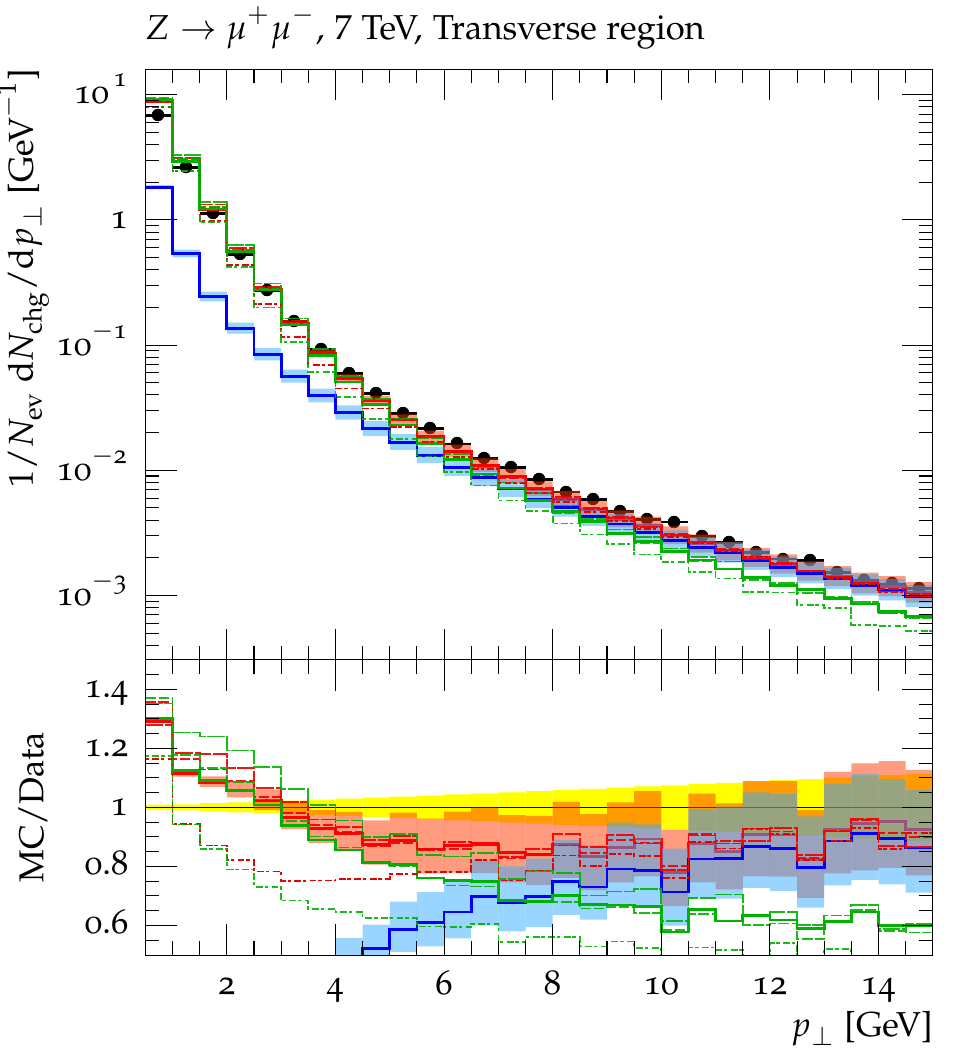}%
\caption{\label{fig:CMS_dNchg_dSumPT}
The differential charged particle multiplicity (top) and transverse momentum (bottom) distributions in the toward region (left panels) and transverse region (right panels).}
\end{center}
\end{figure*}

The \cms analysis uses similar UE-sensitive observables and event region definitions. \cms candidate events also only include di-muon events, with $81 < m_{\mu^{+} \mu^{-}} < 101$ GeV. \Fig{CMS_Nchg_SumPT} shows the \dNchgdetadphi and \dpTsumdetadphi distributions as a function of \ZpT for relatively small $\ZpT < 100$ GeV. As expected, MPI effects are required for a proper description of the data. Again \geneva agrees well with standalone \pythia for both observables. \Fig{CMS_dNchg_dSumPT} shows the differential \Nchg and \pt spectra in the toward and transverse regions, with no restrictions on \ZpT (note that unlike \atlas, here the \pt spectrum is plotted rather than the $\sum\pt$ spectrum). \geneva agrees well with \pythia on the \Nchg distribution, and both underestimate the charged track density for low \Nchg and overestimate it for $\Nchg > 10$. Agreement between \pythia and \geneva on the \pt spectrum holds only for low values of \pt. \geneva agrees with the data within $20\%$ across the whole \pt spectrum, while \pythia predicts a comparatively softer spectrum for $\pt > 3 \GeV$. Again, a more careful theoretical study is required to understand whether the increased agreement can be attributed to \geneva's higher perturbative accuracy. 

\section{Beam thrust event shape}
\label{sec:Beamthrust}

One issue with tuning the modeling of the underlying event is that the observables used in determining the tunes are typically sensitive to several physical effects, both perturbative and nonperturbative. As discussed in the previous section, for UE-sensitive distributions that have dependence on the hard kinematics, the inclusion of the correct perturbative physics is important to describe the data. However, while any higher-order perturbative calculations matched to parton showers can reproduce the hard physics at fixed order, usually they do not account for soft perturbative physics associated with the primary interaction. For this reason, it is difficult to disentangle the perturbative effects of the primary interaction, MPI, and nonperturbative physics when tuning the underlying-event model.

By choosing an observable for which the underlying primary perturbative physics is known precisely, one can get a better handle on the effects due to MPI. Recently, \atlas has measured the normalized beam thrust distribution $1 / \sigma (\df \sigma/\df {\cal \Tau_{\rm CM}})$~\cite{Aad:2016ria}, where
\begin{align}
\Tau_{\rm CM} = \sum_i p_{T,i}\ e^{-|\eta_i|}
\,.\end{align}
Here $p_{Ti}$ and $\eta_i$ are the transverse momentum and rapidity of each particle in the final state but excluding
the decay products of the vector boson.

The beam thrust $0$-jet resolution variable used by \geneva is defined as
\begin{align}
{\Tau_0} = \sum_i p_{T,i}\ e^{-|\eta_i-Y_V|}
\,,\end{align}
where $Y_V$ is the rapidity of the vector boson.
As discussed in \subsec{ShowerMPI}, \geneva includes the perturbative contributions to beam thrust from the primary interaction to $\NNLL'+$NNLO accuracy, which includes in particular soft ISR effects.
While the two observables are not exactly the same, they are closely related and have the same underlying resummation structure~\cite{Stewart:2009yx,Berger:2010xi}. They only differ in the dependence on $Y_V$, leading to some differences in the resummed contributions. However, upon integrating over $Y_V$ and matching to full fixed order, the final distributions for both variables are nearly identical. (A detailed study of this $Y_V$ dependence in a slightly different context can be found in ref.~\cite{Gangal:2014qda}.)
Hence, \geneva essentially predicts the primary perturbative contributions for $\Tau_{\rm CM}$ at NNLL$'+$ NNLO accuracy.

On the other hand, MPI and nonperturbative hadronization effects are not included in \geneva's perturbative input, but have a large effect on the beam thrust spectrum.
Due to the sum over all particles, any secondary collision contributes to beam thrust, such that a prediction without MPI effects fails to describe the data. Also, the experimentally measured distribution is defined by summing only over charged final-state particles, and is thus directly sensitive to hadronization effects.

\geneva matched to \pythiaEight provides the only theoretical calculation of beam thrust, which simultaneously includes $\NNLL'+$NNLO$_0$ logarithmic resummation at low $\Tau_0$, NLO$_1$  accuracy at large $\Tau_0$, as well as the effects from MPI and hadronization. Thus, comparing the predictions of \geneva{}+\pythiaEight to the \atlas measurements allows one to constrain the MPI and nonperturbative effects independent of perturbative contamination.

\begin{figure}[t!]
\begin{center}
\includegraphics[width=\columnwidth]{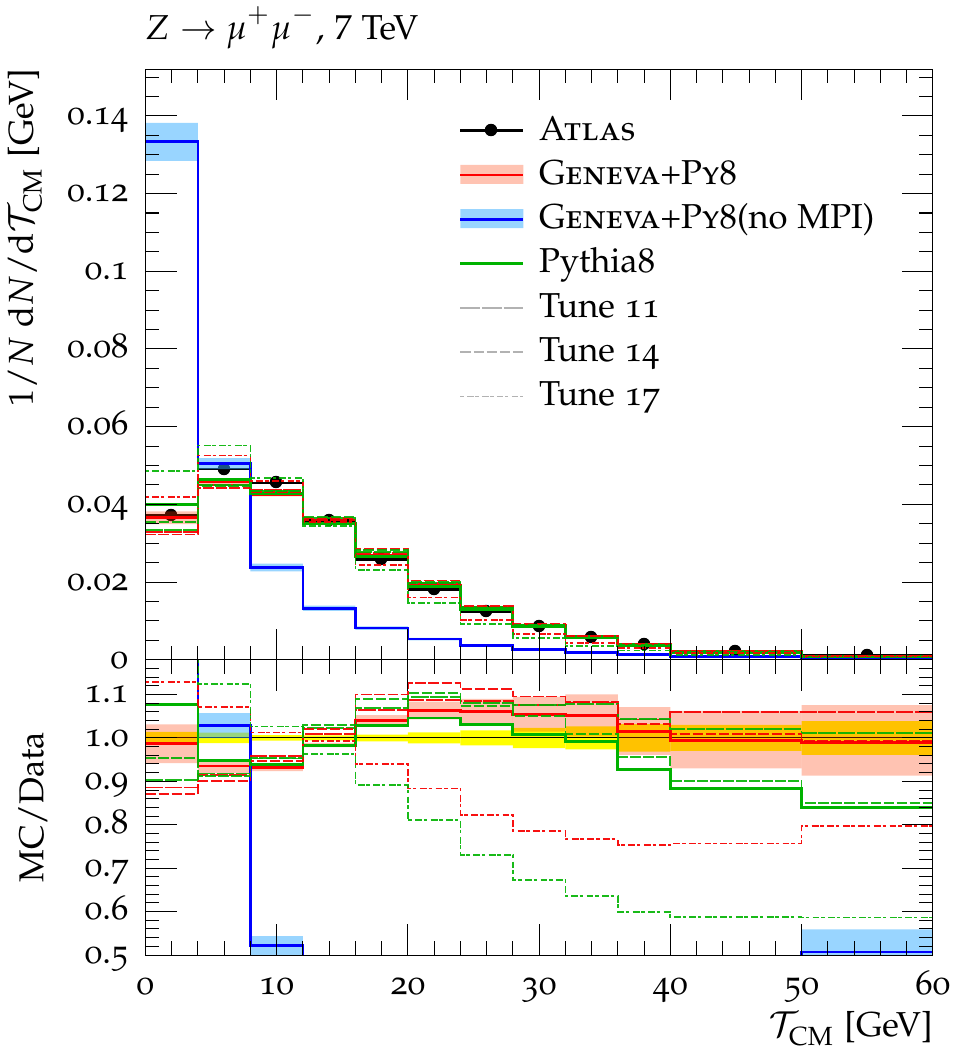}%
\vspace{-2ex}
\caption{The beam thrust distribution $\Tau_{\rm CM}$.}
\label{fig:BeamthrustMain}
\end{center}
\end{figure}

In \fig{BeamthrustMain}, the comparison of \geneva{}+\pythiaEight with the data is shown. One can clearly see that without including the effects of MPI, one cannot reproduce the data. However, once MPI effects are included, \geneva{}+\pythiaEight agrees well with the data. The noticeable exception is when the \atlas AZ tune (tune 17) is used, in which case \pythiaEight standalone and to a lesser extent \geneva{}+\pythiaEight undershoot the data even for moderate values of $ \Tau_{\rm CM}$. This can again be traced back to the lack of recent UE-sensitive inputs in the \atlas AZ tune. Note that while standalone \pythia gives a good agreement with the data for $5\, {\rm GeV} < \Tau_{\rm CM} < 40\, {\rm GeV}$, it falls below the data for $ \Tau_{\rm CM} > 40\, {\rm GeV}$. \geneva{}+\pythiaEight, on the other hand, describes the data much better, especially at larger values of $\Tau_{\rm CM}$. Given that \geneva includes the perturbative soft ISR effects at high logarithmic accuracy, the fact that the predictions are in such good agreement with the data indicates that the MPI is modeled well by the \pythia tune we have chosen.

\begin{figure*}[t!]
\includegraphics[width=0.48\textwidth]{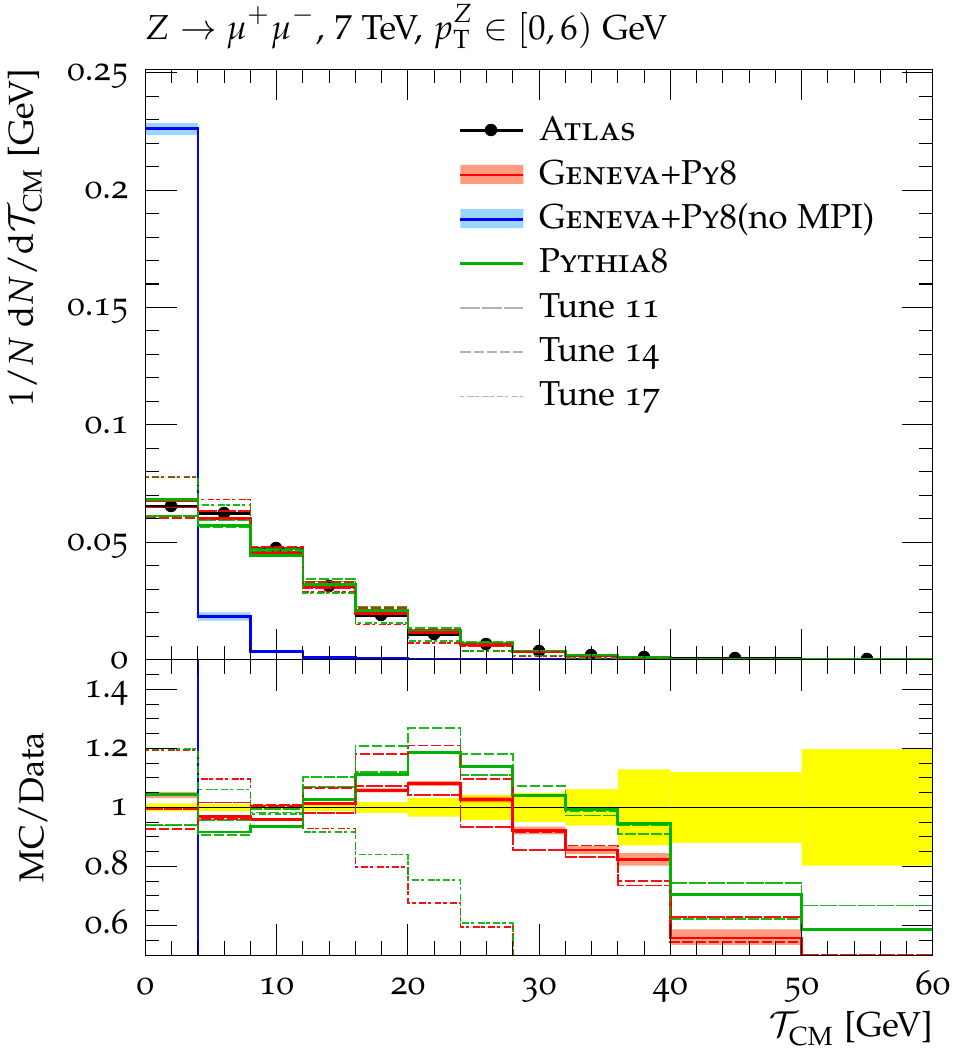}%
\hfill%
\includegraphics[width=0.48\textwidth]{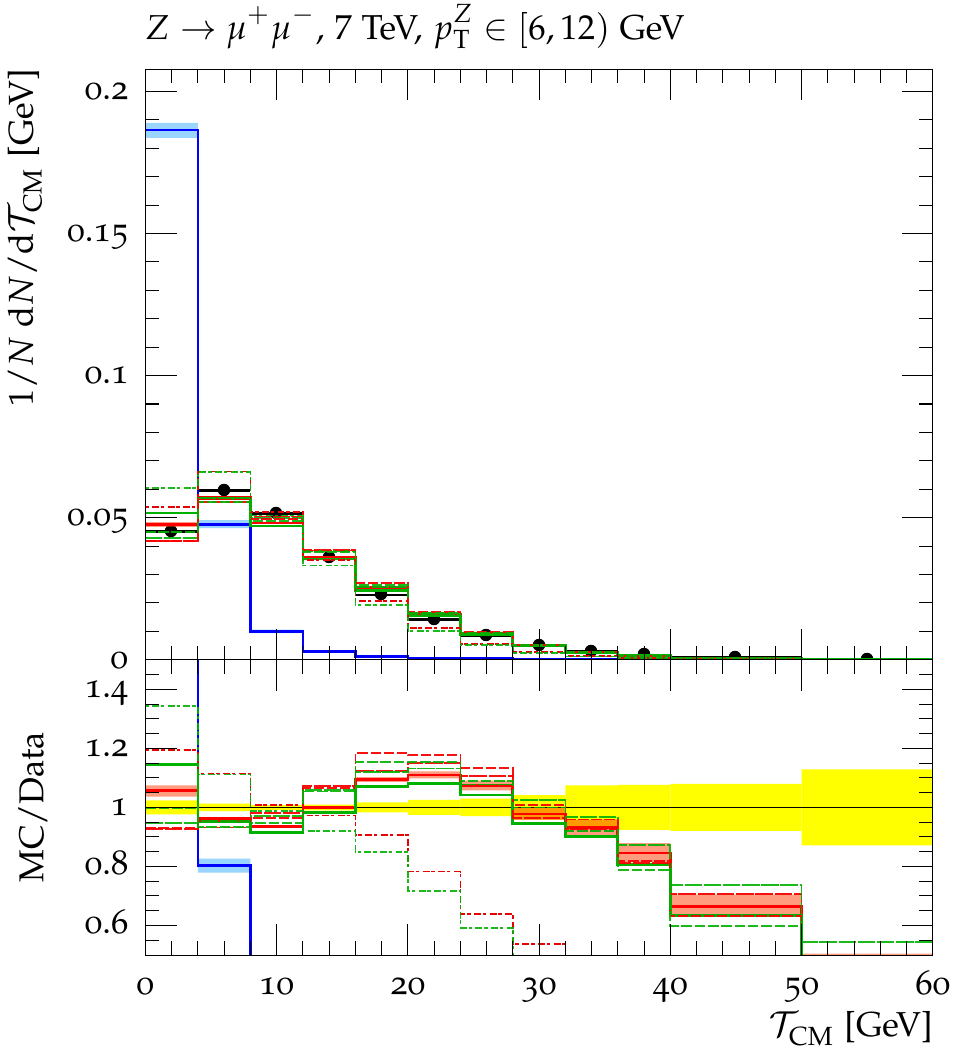}
\\
\includegraphics[width=0.48\textwidth]{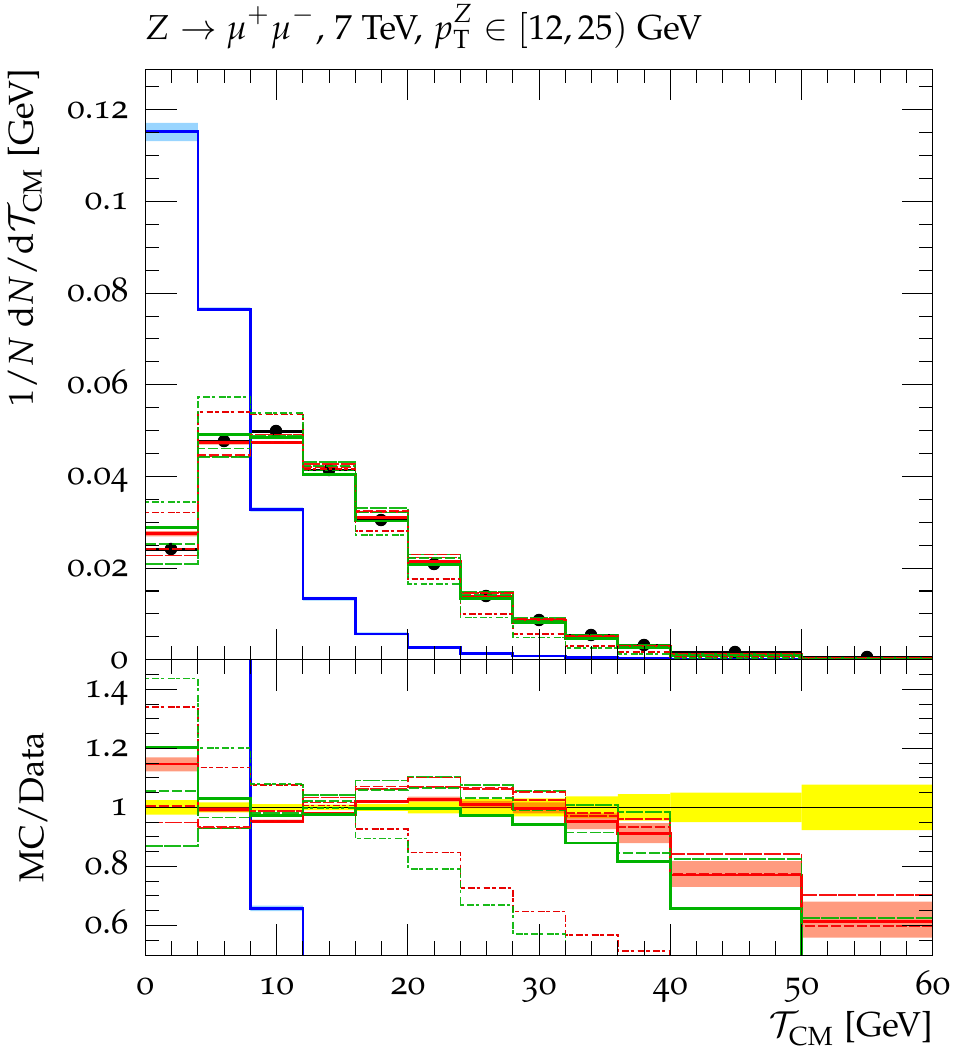}%
\hfill%
\includegraphics[width=0.48\textwidth]{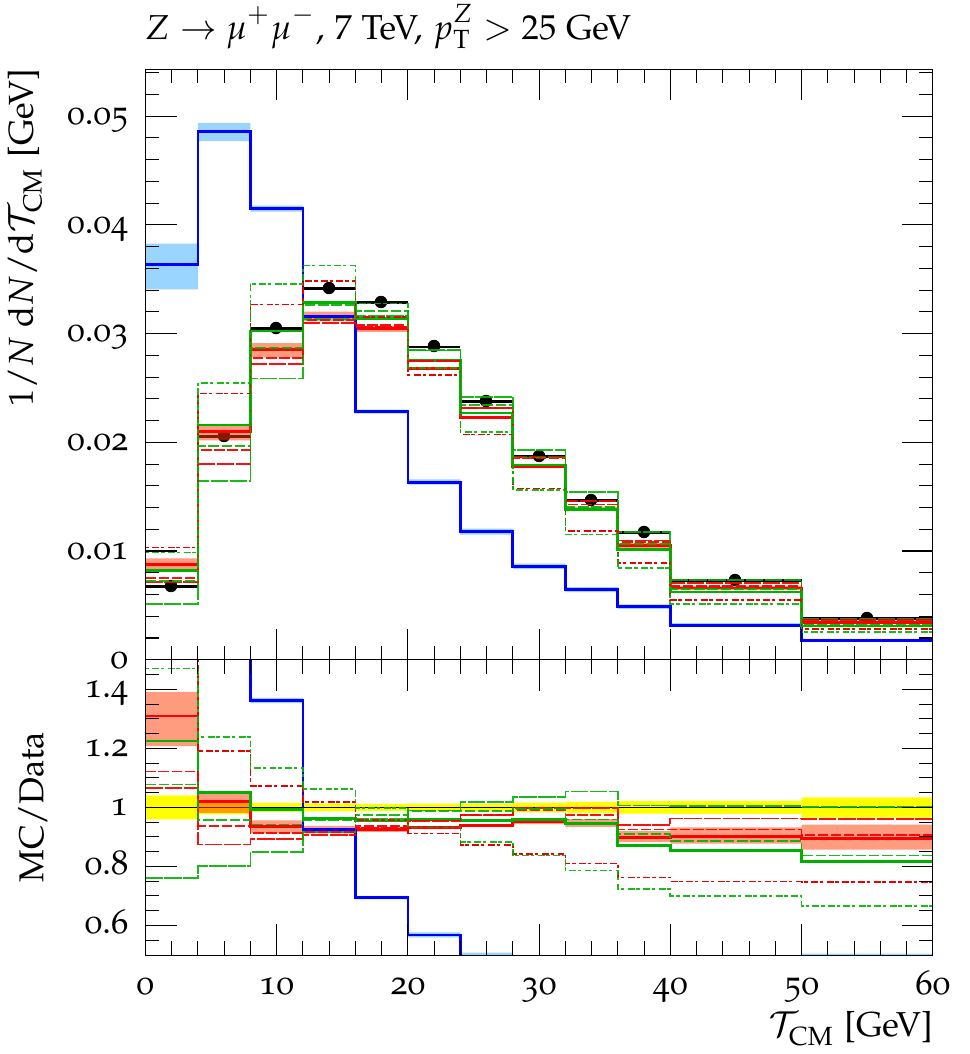}
\caption{The beam thrust distribution $\Tau_{\rm CM}$ for different bins in \ZpT. }
\hspace{-2ex}
\label{fig:BeamthrustPieces}
\end{figure*}

Figure \ref{fig:BeamthrustPieces} compares \geneva{}+\pythiaEight for the $\Tau_{\rm CM}$ distribution in different regions of transverse momentum of the $Z$ boson. This introduces a dependence on the \ZpT spectrum in the measurement. While the overall shape is still described well by \geneva{}+\pythiaEight, a slight discrepancy develops in the tails of the distribution at large $\Tau_{\rm CM}$. This is most likely due to the fact that the \ZpT distribution is predicted with lower accuracy in \geneva compared to the beam thrust distribution. As expected, there is better agreement in the $\ZpT > 25$ range, where the \ZpT spectrum starts to be dominated by the fixed-order calculation.

\section{Conclusions}
\label{sec:conclusions}

We have presented a study of UE-sensitive observables for Drell-Yan neutral-current production in the \geneva Monte Carlo framework.
By adding the ability to turn on the MPI model included in \pythiaEight, one obtains an accurate description of observables that are sensitive to both hard and soft physics. UE-sensitive observables often contain contributions from the primary hard interaction, e.g. charged particle tracks coming from the hadronization of hard jets that recoil against the dilepton system in Drell-Yan processes. In \sec{ComparisonWithData} the predictions of \geneva{}+\pythiaEight are compared against measurements from \atlas and \cms for a variety of UE-sensitive observables. In all cases MPI is clearly needed to accurately describe the data. Using the transverse momentum of the muon pair \ZpT as a measure for the hardness of the event, one finds that the predictions of \geneva agree well with standalone \pythia for events in bins of low \ZpT. This validates that the interface of \geneva with \pythia does not spoil the logarithmic accuracy of the parton shower. For events with high \ZpT, \geneva matches or in some cases improves on the description of standalone \pythia. All comparisons with data are shown for different underlying-event tunes, which are presets of nonperturbative parameters fitted from experimental measurements. The uncertainty coming from the choice of tune is in many cases significant.

The dependence on the tune used in \pythia illustrates the necessity to tune the models contained in \pythia. Traditionally, one uses a wide variety of UE sensitive observables to constrain these models, in particular the effects from MPI. However, as discussed, UE-sensitive observables are not only affected by the details of the MPI model chosen, but also depend on the soft radiation pattern in the primary interaction. For this reason, it is quite difficult to disentangle these two effects from one another. Comparing measurements and predictions  of an observable for which the perturbative effects from the primary interaction is known to higher order, including the soft radiation, will allow to isolate the MPI effects. One such variable is beam thrust, which is predicted in \geneva perturbatively to $\NNLL'+$NNLO and has recently been measured by \atlas. Comparing the predictions of \geneva{}+\pythiaEight with the data allows to test the MPI model more directly than is possible for other UE-sensitive observables. The good agreement with the data indicates that the MPI model in the MonashStar tune we use as our default describes the physics reasonably well.

\begin{acknowledgements}
We thank J.~Lindert and P.~Maierhoefer for help and support in interfacing \geneva to \textsc{OpenLoops}.
CWB would like the thank the KITP in Santa Barbara for financial support during the final stages of this project. 
This work was supported by the Director,
Office of Science, Office of High Energy Physics of the
U.S. Department of Energy under the Contract No. DE-AC02-05CH11231
(CWB, SG), the
DFG Emmy-Noether Grant No. TA 867/1-1 (FT), the COFUND Fellowship
under grant agreement PCOFUND-GA-2012-600377 (SA), 
and a fellowship from the Belgian American Educational Foundation (SG). 
This research used resources of the National Energy Research Scientific Computing Center, which is supported by the Office of Science of the U.S. Department of Energy under Contract No. DE-AC02-05CH11231.

\end{acknowledgements}

\bibliographystyle{../jhep}
\bibliography{../geneva}

\end{document}